\documentclass[ALICE,manyauthors]{cernphprep}
\def\pt{\mbox{$p_{\rm T}$ }}   
\def\pbpb {Pb--Pb\xspace}

\def\jpsi {\mbox{J/$\psi$ }}

\newcommand{\snn}  {\ensuremath{\sqrt{s_{\rm NN}}}}   

\usepackage{datetime}
\usepackage{amsmath}
\usepackage{xcolor}
\usepackage[comma,square,numbers,sort&compress]{natbib}
\usepackage{hyperref}
\usepackage{lineno}
\usepackage{multirow}

\begin{document}%

\begin{titlepage}
\PHyear{2016}
\PHnumber{162}      
\PHdate{24 June}  

\title{J/$\psi$ suppression at forward rapidity in Pb--Pb collisions \\ at  $\mathbf{\sqrt{s_{{\rm NN}}} = 5.02}$~TeV}
\ShortTitle{J/$\psi$ suppression at forward rapidity in Pb--Pb collisions at  $\sqrt{s_{\rm NN}} = 5.02$~TeV}   

\Collaboration{ALICE Collaboration\thanks{See Appendix~\ref{app:collab} for the list of collaboration members}}
\ShortAuthor{ALICE Collaboration} 


\begin{abstract}
The inclusive \jpsi production has been studied in \pbpb and pp collisions at the centre-of-mass energy per nucleon pair $\sqrt{s_{\rm NN}}=5.02$ TeV, using the ALICE detector at the CERN LHC. 
The J/$\psi$ meson is reconstructed, in the centre-of-mass rapidity interval $2.5<y<4$ and in the transverse-momentum range $p_{\rm T}<12$ GeV/$c$, via its decay to a muon pair. 
In this Letter, we present results on the inclusive J/$\psi$ cross section in pp collisions at $\sqrt{s}=5.02$ TeV and on the nuclear modification factor $R_{\rm AA}$. The latter is presented as a function of the centrality of the collision and, for central collisions, as a function of the transverse momentum $p_{\rm T}$ of the J/$\psi$. 
The measured $R_{\rm AA}$ values indicate a suppression of the J/$\psi$ in nuclear collisions and are then compared to our previous results obtained in \pbpb collisions at $\sqrt{s_{\rm NN}}=2.76$~TeV. 
The ratio of the $R_{\rm AA}$ values at the two energies is also computed and compared to calculations of statistical and dynamical models. 
The numerical value of the ratio for central events (0--10\% centrality) is 
$1.17 \pm 0.04 {\rm{(stat)}}\pm 0.20 {\rm{(syst)}}$. 
In central events, as a function of $p_{\rm T}$, a slight increase of $R_{\rm AA}$ with collision energy is visible in the region $2<p_{\rm T}<6$ GeV/$c$. Theoretical calculations qualitatively describe the measurements, within uncertainties.
\end{abstract}
\end{titlepage}
\setcounter{page}{2}

%
%

\section{Introduction}

When heavy nuclei collide at ultrarelativistic energies, a state of strongly-interacting matter is formed, characterised by high temperature and density, where quarks and gluons are not confined into hadrons (Quark--Gluon Plasma, QGP~\cite{Shuryak:1978ij}).  
A detailed characterisation of the QGP is the object, since more than 25 years, of an intense research activity at the CERN/SPS~\cite{Heinz:2000bk} and at the BNL/RHIC~\cite{Arsene:2004fa,Back:2004je,Adams:2005dq,Adcox:2004mh} and CERN/LHC~\cite{Muller:2012zq} ion colliders. 
Charmonia and bottomonia, which are bound states of charm-anticharm (${\rm c{\overline c}}$) or bottom-antibottom (${\rm b{\overline b}}$) quarks, respectively~\cite{Brambilla:2010cs}, are among the most sensitive probes of the characteristics of the QGP. 
A suppression of their yields in nucleus--nucleus (A--A) collisions with respect to expectations from proton--proton (pp) collisions was experimentally observed. 
For the J/$\psi$ meson, the ground ${\rm c{\overline c}}$ state with quantum numbers J$^{\rm PC}=1^{--}$, a suppression was found at the SPS, in Pb--Pb and In--In interactions at the centre-of-mass energy per nucleon pair $\sqrt{s_{\rm NN}}=17.2$ GeV~\cite{Alessandro:2004ap,Arnaldi:2007zz}, RHIC, in Au--Au  interactions at $\sqrt{s_{\rm NN}}=200$ GeV~\cite{Adare:2011yf,Abelev:2009qaa},  and finally at the LHC, in \pbpb collisions at $\sqrt{s_{\rm NN}}=2.76$ TeV~\cite{Abelev:2012rv,Chatrchyan:2012np}. 
Early theoretical calculations predicted J/$\psi$ suppression to be induced by the screening of the colour force in a deconfined medium and to become stronger as the QGP temperature increases~\cite{Matsui:1986dk,Digal:2001ue}. 
In a complementary way to this static approach, J/$\psi$ suppression can also be seen as the result of dynamical interactions with the surrounding partons~\cite{Ferreiro:2012rq,Zhao:2011cv,Zhou:2014kka}.
The LHC results, integrated over transverse momentum ($p_{\rm T}$) down to $p_{\rm T}=0$, show a suppression of the J/$\psi$, quantified through the ratio between its yields in \pbpb and those in pp, normalised to the number of nucleon--nucleon collisions in \pbpb (nuclear modification factor, $R_{\rm AA}$). However, the observed suppression is smaller than at SPS and RHIC~\cite{Abelev:2013ila,Adam:2015isa}, in spite of the higher initial temperature of the QGP formed at the LHC~\cite{Adam:2015lda}. The effect is particularly evident for head-on (central) collisions. In order to explain these observations, theoretical models require a contribution from J/$\psi$ regeneration via a recombination mechanism~\cite{BraunMunzinger:2000px,Thews:2000rj} between the ${\rm c}$ and ${\rm \overline c}$ quarks, during the deconfined phase and/or at the hadronisation of the system, which occurs when its temperature falls below the critical value $T_{\rm c}\sim 155$ MeV~\cite{Bhattacharya:2014ara}. The strength of this regeneration effect increases with the initial number of produced ${\rm c{\overline c}}$ pairs relative to the total number of quarks and, therefore, increases with the collision energy, explaining the reduced suppression at the LHC. Since the bulk of charm-quark production occurs at small momenta, recombination should be more important for low-$p_{\rm T}$ J/$\psi$, as observed in the LHC results~\cite{Adam:2015isa}.

An important test of the suppression and regeneration picture of J/$\psi$ production at the LHC can be obtained by comparing the centrality and $p_{\rm T}$ dependence of the J/$\psi$ $R_{\rm AA}$, measured at $\sqrt{s_{\rm NN}}=2.76$ TeV, to that obtained at $\sqrt{s_{\rm NN}}=5.02$ TeV, the highest energy available up to now in nuclear collisions. 
The suppression effects related to colour screening should become stronger when increasing the collision energy, due to the higher QGP temperature, and also the recombination effects should become stronger, due to the expected increase of the $\rm c\overline c$ production cross section. The two effects act in opposite directions and the 
comparison of the $R_{\rm AA}$ at the different energies can provide insights in the evolution of the relative contribution of the two processes.  

In this Letter, we present the first results on the J/$\psi$ $R_{\rm AA}$ measured by the ALICE Collaboration in \pbpb collisions at $\sqrt{s_{\rm NN}}=5.02$ TeV and the integrated and $p_{\rm T}$ differential J/$\psi$ production cross section in pp collisions at the same energy.
In both \pbpb and pp collisions, the J/$\psi$ is reconstructed via its dimuon decay channel at forward rapidity, $2.5<y<4$ and for $p_{\rm T}<12$ GeV/$c$. The measurements refer to inclusive J/$\psi$ production, that includes both prompt J/$\psi$ (direct J/$\psi$ and feed-down from higher-mass resonances) and non-prompt J/$\psi$ (from decay of beauty hadrons). The nuclear modification factor is obtained by normalising the J/$\psi$ yield in \pbpb collisions to the product of the nuclear overlap function times the corresponding J/$\psi$ cross section measured in pp, at the same energy and in the same kinematic window.
The results on $R_{\rm AA}$ are presented as a function of the J/$\psi$ $p_{\rm T}$ and of the centrality of the collision.

\section{Experimental apparatus and data sample}
\label{section:Apparatus}

The ALICE detector design and performance are extensively described in~\cite{Aamodt:2008zz} and~\cite{Abelev:2014ffa}. 
The analysis presented here is based on the detection of muons in the forward muon spectrometer~\cite{Aamodt:2011gj}, which covers the pseudo-rapidity range $-4<\eta<-2.5$~\footnote{In the ALICE reference frame, the muon spectrometer covers a negative $\eta$ range and consequently a negative $y$ range. We have chosen to present our results with a positive $y$ notation.}. 
In addition, the Silicon Pixel Detector (SPD)~\cite{Aamodt:2010aa} is used to reconstruct the primary vertex. The V0 detectors~\cite{Abbas:2013taa} provide a minimum-bias (MB) trigger and are used to determine the centrality of the collision, while the T0 Cherenkov counters~\cite{Bondila:2005xy} are used for the luminosity determination in pp collisions. Finally, the Zero Degree Calorimeters (ZDC) are used to reject electromagnetic \pbpb interactions~\cite{ALICE:2012aa}. A brief description of these detectors is given hereafter.

The muon spectrometer contains a front absorber, made of carbon, concrete and steel,  placed between 0.9 and 5 m from the Interaction Point (IP), which filters out hadrons, thus decreasing the occupancy in the downstream tracking system. The latter is  composed of five stations, each one consisting of two planes of Cathode Pad Chambers (CPC). The third tracking station is placed inside the gap of a dipole magnet with a 3~T$\cdot$m field integral. 
Two trigger stations, each one equipped with two planes of Resistive Plate Chambers (RPC), are located behind a 7.2 interaction length iron wall, which absorbs secondary hadrons escaping the front absorber and low-momentum muons. 
The muon trigger system delivers single-muon and dimuon triggers with a programmable transverse-momentum threshold.
Finally, throughout its entire length, a conical absorber around the beam pipe ($\theta<2^{\circ}$) made of tungsten, lead and steel shields the muon spectrometer against secondary particles produced by the interaction of large-$\eta$ primary particles in the beam pipe.

The primary vertex is reconstructed using hit pairs in the two cylindrical layers of the SPD~\cite{Aamodt:2008zz,Aamodt:2010aa}, which have average radii of 3.9 and 7.6 cm, and cover the pseudo-rapidity intervals $|\eta|<2$ and $|\eta|<1.4$, respectively. 

The two V0 detectors~\cite{Abbas:2013taa}, with 32 scintillator tiles each, are placed on each side of the IP, covering the pseudo-rapidity ranges $2.8<\eta< 5.1$ and $-3.7<\eta<-1.7$. 
The coincidence of the signals from the two hodoscopes defines the MB trigger. Beam-induced background is reduced by applying timing cuts on the signals from the V0s and ZDCs. The latter are positioned along the beam direction at $\pm$112.5~m from the IP. 
Finally, the T0 detectors~\cite{Bondila:2005xy}, made of two arrays of quartz Cherenkov counters, are placed on both sides of the IP, covering the pseudo-rapidity intervals $-3.3<\eta< -3$ and $4.6<\eta<4.9$. 

In \pbpb collisions, the centrality determination is based on a Glauber fit of the total V0 signal amplitude distribution as described in~\cite{Abelev:2013qoq,Adam:2015ptt}.
 A selection corresponding to the most central 90\% of the hadronic cross section was applied; for these events the MB trigger is fully efficient.
 
For both \pbpb and pp data taking, the trigger condition used in the analysis is a $\mu\mu$-MB trigger formed by the coincidence of the MB trigger and an unlike-sign (US) dimuon trigger. The latter has a trigger probability for each of the two muon candidates that increases with the muon $p_{\rm T}$, is 50\% at 1.0~GeV/$c$ (0.5 GeV/$c$) in \pbpb (pp) collisions, and saturates at $p_{\rm T}\approx 2.5$ GeV/$c$, where it reaches a value of about 98\%.  
Like-sign dimuon triggers were also collected, mainly for background normalisation purposes in the \pbpb\ analysis.

The data samples used in this analysis correspond to an integrated luminosity $L_{\rm int}^{\text {Pb-Pb}}\approx 225\, \mu{\rm b}^{-1}$ for \pbpb\ and 
$L_{\rm int}^{\text {pp}}\approx 106\, {\rm nb}^{-1}$ for pp collisions.

\section{Data analysis}\label{section:analysis}

The analysis procedure was very similar for the two data samples described in this Letter. In the following paragraphs, the \pbpb\ analysis is first presented, followed by the description of the pp one. 

The \jpsi candidates were formed by combining pairs of US tracks reconstructed in the geometrical acceptance of the muon spectrometer using the tracking algorithm described in~\cite{Aamodt:2011gj}. 
The same single-muon and dimuon selection criteria as in previous analyses~\cite{Adam:2015isa} were applied,  and tracks in the tracking system were required to match a track segment in the muon trigger system (trigger tracklet).

The \jpsi raw yields were determined from the invariant mass distribution of US dimuons using two methods. 
In the first one, the US dimuon invariant mass distributions were fitted with the sum of a signal and a background function. 
In the second approach, the background, estimated using an event-mixing technique and normalised using the like-sign dimuon distributions~\cite{Adam:2015isa}, was subtracted and the resulting spectra were fitted with the sum of a signal function and a (small) residual background component.

Various shapes were considered for the signal and background contributions. 
For the \jpsi signal either an extended Crystall Ball (CB2) function or a pseudo-Gaussian with a mass-dependent width were used~\cite{ALICE-PUBLIC-2015-006}. 
The non-Gaussian tails of the signal functions were fixed either (i) to the values obtained in Monte Carlo (MC) simulations, where simulated $\rm{J}/\psi \rightarrow \mu^{+}\mu^{-}$ are embedded into real events to account for the effect of the detector occupancy, or (ii) to the values obtained in a high-statistics pp collision sample at $\sqrt{s}=13$ TeV, collected under similar detector conditions.
The tail parameters exhibit a dependence on the \pt and rapidity of the \jpsi and a mild dependence on the centrality of the collision. The small contribution of the $\psi(2S)$ signal was taken into account in the fits, its mass and width being tied to those of the J/$\psi$~\cite{Abelev:2014zpa}.
For the background, when the US dimuon mass spectrum was fitted, a  variable-width-Gaussian with a mass-dependent width or the ratio of a 2$^{\rm nd}$ to a 3$^{\rm rd}$ order polynomial were used. 
When considering the US dimuon distributions after subtraction of the background obtained with the event-mixing  procedure, a small dimuon continuum component is still present and was fitted using the sum of two exponentials. 
Several fitting sub-ranges, within the interval $2<m_{\mu\mu}<5$ GeV/$c^2$, were used for both signal extraction procedures.

\begin{figure}[h]
\begin{center}
\includegraphics[width=0.96\linewidth]{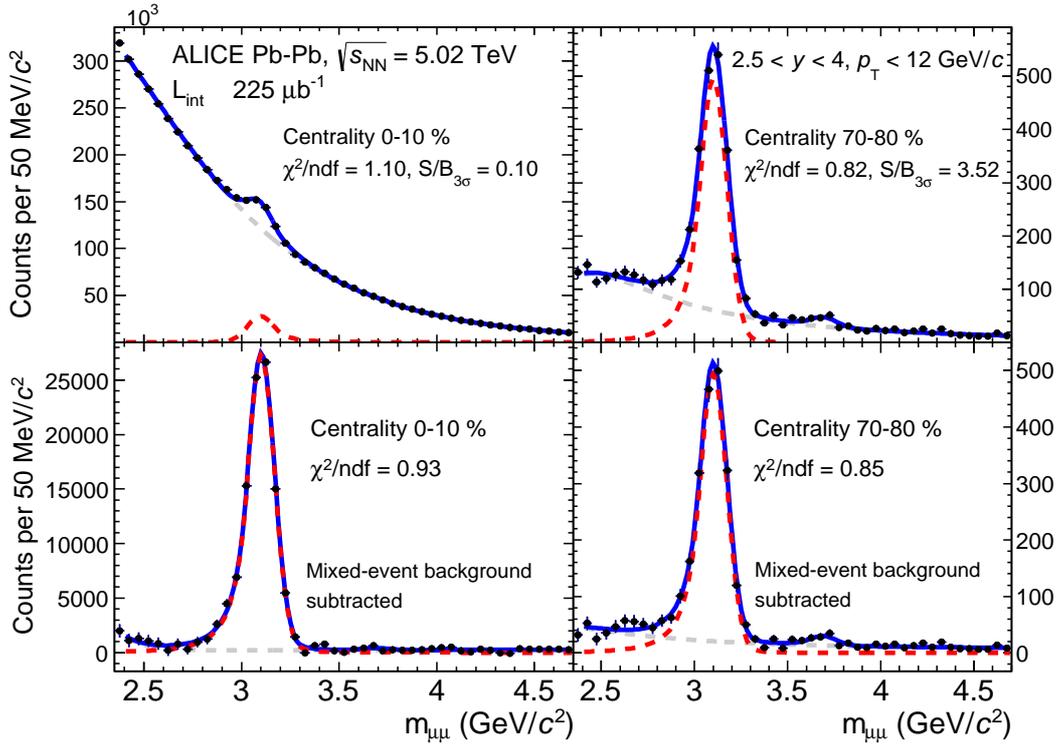} 
\caption{(colour online). Invariant mass distributions of US dimuons with $2.5 < y < 4$ and $p_{\rm T}<12$ GeV/$c$. 
The top (bottom) row shows the distribution before (after) background subtraction with the event-mixing technique. The left panels correspond to the most central events 
(0--10\%) while the right panels to a peripheral (70--80\%) centrality range. 
The fit curves shown in blue represent the sum of the signal and background shapes, while the red lines correspond to the J/$\psi$ signal and the grey ones to the background.}
\label{fig:1}
\end{center}
\end{figure}

Figure~\ref{fig:1} shows examples of fits to the US dimuon invariant mass distributions with and without background subtraction using the event-mixing technique, for different selections in centrality. 
The raw \jpsi yield in each centrality or \pt interval was determined as the average of the results obtained with the two fitting approaches, the various parametrisations of signal and background and the different fitting ranges, while the corresponding systematic uncertainties were defined as the RMS of these results.
A further contribution to the systematic uncertainty was estimated by using a different set of resonance tails obtained using in the MC simulation a different particle transport model (GEANT4~\cite{Agostinelli:2002hh} instead of GEANT3~\cite{Brun:1994aa}). 
The total number of  $\rm{J}/\psi$, integrated over centrality, $p_{\rm T}$ and $y$, is   $N_{{\rm J}/\psi} = 2.77 \pm 0.02({\rm stat}) \pm 0.05({\rm syst}) \cdot 10^5$. The systematic uncertainty ranges from 1.6\% to 2.8\% as a function of centrality and from 1.2\% to 3.1\% as a function of $p_{\rm T}$.

The nuclear modification factor, as a function of the centrality class $i$ of the collision and for the J/$\psi$ transverse-momentum interval  $\Delta p_{\rm T}$, is calculated as 
\begin{eqnarray}
R_{\rm{AA}}^{i}(\Delta p_{\rm T}) = \frac{N_{{\rm J}/\psi}^{i}(\Delta p_{\rm T})}{{\rm BR}_{\rm{J/}\psi\rightarrow\mu^{+}\mu^{-}}  N_{\rm{MB}}^{i}  A \epsilon^{i}(\Delta p_{\rm T}) \langle T_{\rm{AA}}^{i} \rangle   \sigma_{\rm{J/}\psi}^{\rm{pp}}(\Delta p_{\rm T})},
\label{eq:raa}
\end{eqnarray}
where $N_{\rm{J/}\psi}^{i}(\Delta p_{\rm T})$ is the number of extracted \jpsi in a given centrality and $p_{\rm T}$ range, ${\rm BR}_{\rm{J/}\psi\rightarrow\mu^{+}\mu^{-}} = 5.96 \pm 0.03$\% is the branching ratio of the dimuon decay channel~\cite{Agashe:2014kda}, $N_{\rm{MB}}^{i}$ is the number of equivalent minimum-bias events, $A\epsilon^{i}(\Delta p_{\rm T})$ is the product of the detector acceptance times the reconstruction efficiency, $\langle T_{\rm{AA}}^{i} \rangle$ is the average of the nuclear overlap function, and $\sigma_{\rm{J/}\psi}^{\rm{pp}}(\Delta p_{\rm T})$ is the inclusive \jpsi cross section for pp collisions at the same energy and in the same kinematic range as the \pbpb data. 

The $A\epsilon$ values were determined from MC simulations, with the generated \pt and $y$ distributions for the \jpsi\ adjusted on data, and separately tuned for each centrality class using an iterative approach. Unpolarised J/$\psi$ production was assumed~\cite{Adam:2015isa}.
For the tracking chambers, the time-dependent status of each electronic channel during the data taking period was taken into account as well as the misalignment of the detection elements. 
The efficiencies of the muon trigger chambers were determined from data and were then applied in the simulations. 
Finally, the dependence of the efficiency on the detector occupancy was taken into account by embedding MC-generated J/$\psi$ into real minimum-bias \pbpb events. 

For \jpsi produced within $2.5 < y < 4$ and $p_{\rm T}<12$ GeV/$c$, in 0--90\% most central collisions, the  $A\epsilon$ value is $0.136 \pm 0.007$(syst). 
A relative decrease of the efficiency by 14\% was observed when going from peripheral to central collisions. As a function of $p_{\rm T}$, $A\epsilon$ has a minimum value of about $0.12$ at $p_{\rm T}\approx 1.5$ GeV/$c$, and then steadily increases up to about 0.4 at the upper end of the considered range. 
The following sources of systematic uncertainty on $A\epsilon$ were considered. 
A first contribution of 2\% due to the input MC \pt and $y$  distributions was estimated by (i) varying the input shapes that were tuned on data within their statistical uncertainties and (ii) taking into account the effect of possible $p_{\rm T}-y$ correlations by comparing, as a function of centrality, the $A\epsilon$ values with the corresponding result of a 2-D acceptance calculation in classes of $p_{\rm T}$ and $y$. 
A second contribution comes from the tracking efficiency and it was estimated by comparing the single-muon tracking efficiency values obtained, in MC and data, with a procedure that exploits the redundancy of the tracking-chamber information~\cite{Adam:2015isa}.
A 3\% systematic uncertainty on the dimuon tracking efficiency is obtained and is approximately constant as a function of centrality and kinematics. 
The systematic uncertainty on the dimuon trigger efficiency represents the third contribution and it has two origins:  the intrinsic efficiencies of the muon trigger chambers and the response of the trigger algorithm. 
The first one was determined from the uncertainties on the trigger chamber efficiencies measured from data and applied to simulations and it amounts to 1.5\%.
The second one was estimated by comparing the $p_{\rm T}$ dependence, at the single-muon level, of the trigger response function between data and MC and it varies between 0.2\% and 4.6\% as a function of $p_{\rm T}$. 
Combining the two sources, a systematic uncertainty ranging from 1.5\% to 4.8\% as a function of the J/$\psi$ $p_{\rm T}$ is obtained.
Finally, there is a 1\% contribution related to the choice of the $\chi^2$ cut used in defining the matching between the reconstructed tracks and the trigger tracklets. 

The normalisation factor to the number of equivalent MB events was obtained as 
$N_{\rm{MB}}^{i}=F^{i}\cdot N_{\mu\mu\textnormal{-MB}}$, where $N_{\mu\mu\textnormal{-MB}}$ is the number of $\mu\mu\textnormal{-MB}$ triggered events, and $F^{i}$ is the inverse of the probability of having a dimuon trigger in a MB event in the centrality range $i$. 
The $F^{i}$ values were calculated with two different methods, by applying the dimuon trigger condition in the analysis on minimum-bias events, or from the relative counting rate of the two triggers~\cite{Abelev:2013yxa}. 
The obtained value, in the 0--90\% centrality class, is $F=11.84\pm 0.06$, where the  uncertainty is dominated by a systematic contribution  corresponding to the difference between the results obtained with the two approaches. 
As a function of centrality, $F^{i}=F\cdot\Delta^{i}$, where $\Delta^{i}$ is the fraction of the inelastic cross section of a given centrality class with respect to the whole 0--90\% centrality range (e.g. 0.1/0.9 for 0--10\% centrality and so on).

The values for $\langle T_{\rm{AA}}^{i} \rangle$ and for the average number of participant nucleons $\langle N_{\rm part}^{i}\rangle$ were obtained via a Glauber calculation~\cite{Abelev:2013qoq,Adam:2015ptt,ALICE-PUBLIC-2015-008}. The systematic uncertainty is 3.2\% for the 0--90\% centrality range and was obtained by varying within uncertainties the density parameters of the Pb nucleus and the nucleon--nucleon inelastic cross section~\cite{Adam:2015ptt,ALICE-PUBLIC-2015-008}.

Finally, the effects of the uncertainty on the value of the V0 signal amplitude corresponding to 90\% of the hadronic \pbpb cross section were estimated by varying such a value by $\pm0.5$\%~\cite{Abelev:2013qoq} and redefining correspondingly the centrality intervals. The systematic effect on $R_{\rm AA}$ ranges from 0.1\% to 6.6\% from central to peripheral collisions.

The J/$\psi$ cross-section values in pp collisions at $\sqrt{s}=5.02$ TeV, both integrated and $p_{\rm T}$ differential, were obtained with an analysis procedure similar to the one described in the previous paragraphs for Pb--Pb.
In particular, the same criteria for single-muon and dimuon selection were adopted. 

The signal extraction was then performed by fitting the spectra with the sum of a signal and a background contribution, using shapes similar to those adopted for the \pbpb\ analysis. The background subtraction via the event-mixing technique was not used, as the signal-over-background ratio is larger by a factor $\sim 40$, in the $p_{\rm T}$-integrated spectra, with respect to central \pbpb collisions,  making the influence of the background estimate much less important in the determination of the uncertainty on $N^{\rm pp}_{{\rm J}/\psi}$. The value $N^{\rm pp}_{{\rm J}/\psi} =8649 \pm 123 {\rm {(stat)}}\pm 297 {\rm {(syst)}}$ is obtained, with the systematic uncertainty determined as for the \pbpb\ analysis.

The determination of $A\epsilon_{\rm pp}$ was carried out via MC simulations. Since no appreciable dependence of the tracking efficiency as a function of the hadronic multiplicity can be seen in pp, a pure MC (i.e., without embedding) was used. The input $p_{\rm T}$ and $y$ distributions were obtained from the measured ones via an iterative procedure, and unpolarised J/$\psi$ production was assumed~\cite{Abelev:2011md}. The obtained value is $A\epsilon_{\rm pp} = 0.243 \pm 0.007 ({\mathrm {syst}})$,  with the systematic uncertainties on the tracking, trigger and matching efficiency calculated as in the \pbpb\ analysis. Because of the limited pp statistics, the systematic uncertainty on the MC inputs was not obtained through a 2-D acceptance calculation, as done in the \pbpb\ analysis, but it was determined comparing the $A\epsilon$ values obtained using J/$\psi$ 
$p_{\rm T}$ ($y$) distributions evaluated in various $y$ ($p_{\rm T}$) intervals in pp collisions at $\sqrt{s}=7$ TeV~\cite{Aaij:2011jh}.

The integrated luminosity was calculated as $L_{\rm int}^{\rm pp}= (N_{\mu\mu\textnormal{-MB}}^{\rm pp}\cdot F^{\rm pp})/\sigma_{\rm ref}^{\rm pp}$, where $\sigma_{\rm ref}^{\rm pp}$ is a reference-trigger cross section measured in a van der Meer scan, following the procedure detailed in~\cite{ALICE-PUBLIC-2016-005}, and $F^{\rm pp}$ is the ratio of the reference-trigger probability to the $\mu\mu$-MB trigger probability. The corresponding numerical value is $L_{\rm int}^{\rm pp}= 106.3 \pm 2.2 {(\rm {syst})}$ nb$^{-1}$, where the quoted uncertainty reflects the van der Meer scan uncertainty.

Finally, the inclusive J/$\psi$ cross section in pp collisions at $\sqrt{s}=5.02$ TeV was obtained as 
\begin{equation}
{\frac{{\rm d}^2\sigma_{\rm{J/}\psi}^{\rm{pp}}}{{\rm d}y{\rm d}p_{\rm T}}} = 
{\frac{N^{\rm pp}_{{\rm J}/\psi}(\Delta p_{\rm T})}{{\rm BR}_{\rm{J/}\psi\rightarrow\mu^{+}\mu^{-}}\,
L_{\rm int}^{\rm pp}\, A \epsilon_{\rm pp}(\Delta p_{\rm T})\Delta p_{\rm T}\Delta y}} .
\label{eq:sigma}
\end{equation}

Table 1 summarises the systematic uncertainties on the measurement of the nuclear modification factors and ${\rm d}^2\sigma^{\rm pp}_{\rm J/\psi}/{\rm d}y{\rm d}p_{\rm T}$. 

The $R_{\rm AA}$ values presented in the following refer to inclusive J/$\psi$ production, i.e. include both prompt and non-prompt J/$\psi$.  
Since beauty-hadron decays occur outside the QGP, the non-prompt J/$\psi$ $R_{\rm AA}$ is related to the nuclear modification of the beauty-hadron $p_{\rm T}$ distributions. 
The difference between the $R_{\rm AA}$ of prompt and inclusive J/$\psi$ can be estimated as in~\cite{Adam:2015isa},  using the fraction $F_{\rm B}$ of non-prompt to inclusive J/$\psi$ in pp collisions and assuming two extreme cases for the $R_{\rm AA}^{\text {non-prompt}}$ of non-prompt J/$\psi$, namely no medium effects on b-quarks ($R_{\rm AA}^{\text {non-prompt}}=1$) or their complete suppression ($R_{\rm AA}^{\text {non-prompt}}=0$). 
$F_{\rm B}$ was obtained by an interpolation of the LHCb measurements in pp collisions at $\sqrt{s}=2.76$ and 7 TeV~\cite{ALICE:2013spa,Aaij:2011jh,Aaij:2012asz}. The quantitative effect on the inclusive J/$\psi$ $R_{\rm AA}$ is provided in the following along with the results.

\begin{table}
\begin{center}
\begin{tabular}{c|c|c|c|c|c}
& \multicolumn{3}{c|}{$R_{\rm AA}$} &  \multicolumn{2}{|c}{${\rm d}^2\sigma^{\rm pp}_{{\rm J}/\psi}/{\rm d}y{\rm d}p_{\rm T}$} \\
\hline
Source & 0--90\% & vs $p_{\rm T}$ & vs centrality \\
 & $p_{\rm T}<12$ GeV/$c$ & (0--20\%) & ($p_{\rm T}<8$ GeV/$c$) & $p_{\rm T}<12$ GeV/$c$ & vs $p_{\rm T}$ \\ 
\hline
\hline
Signal extr. & 1.8 & 1.2--3.1 & 1.6--2.8 & 3 & 1.5--9.3 \\
\hline
MC input & 2 & 2 & 2$^*$ & 2 & 0.7--1.5 \\
\hline
Tracking eff. & 3 & 3 & 3$^*$ & 1 & 1 \\
\hline
Trigger eff. & 3.6 & 1.5--4.8 & 3.6$^*$ & 1.8 & 1.5--1.8 \\
\hline
Matching eff. & 1 & 1 & 1$^*$ & 1 & 1 \\
\hline
$F$ ($L^{\rm pp}_{\rm int}$) & 0.5 & 0.5$^*$ & 0.5$^*$ & (2.1) & (2.1$^*$) \\
\hline
BR & - & - & - & 0.5 & 0.5*\\
\hline
$\langle T_{\rm AA}\rangle$ & 3.2 & 3.2$^*$ & 3.1--7.6 & \multicolumn{2}{c}{\multirow{3}{*}{}}\\
\cline{1-4}
Centrality & 0 & 0.1$^*$ & 0--6.6 & \multicolumn{2}{c}{}\\
\cline{1-4}
pp reference & 5.0 & 3--10 $\bigoplus$ 2.1$^*$($L^{\rm pp}_{\rm int}$) & 4.9$^*$ & \multicolumn{2}{c}{}\\
\cline{1-4}
\end{tabular}
\caption{Summary of systematic uncertainties, in percentage, on $R_{\rm AA}$ and ${\rm d}^2\sigma^{\rm pp}_{{\rm J}/\psi}/{\rm d}y{\rm d}p_{\rm T}$. Values marked with an asterisk correspond to correlated uncertainties as a function of $p_{\rm T}$ (second and fifth column) or centrality (third column). There is no correlation between the uncertainties related to the analysis of the \pbpb\ and of the pp sample. The contents of the ``pp reference'' row correspond to the quadratic sum of the contributions indicated for ${\rm d}^2\sigma^{\rm pp}_{{\rm J}/\psi}/{\rm d}y{\rm d}p_{\rm T}$, excluding only the BR uncertainty which cancels out when forming the $R_{\rm AA}$.}
\end{center}
\label{tab:1}
\end{table}

\section{Results}\label{section:results}

\begin{figure}
\begin{center}
\includegraphics[width=0.7\linewidth]{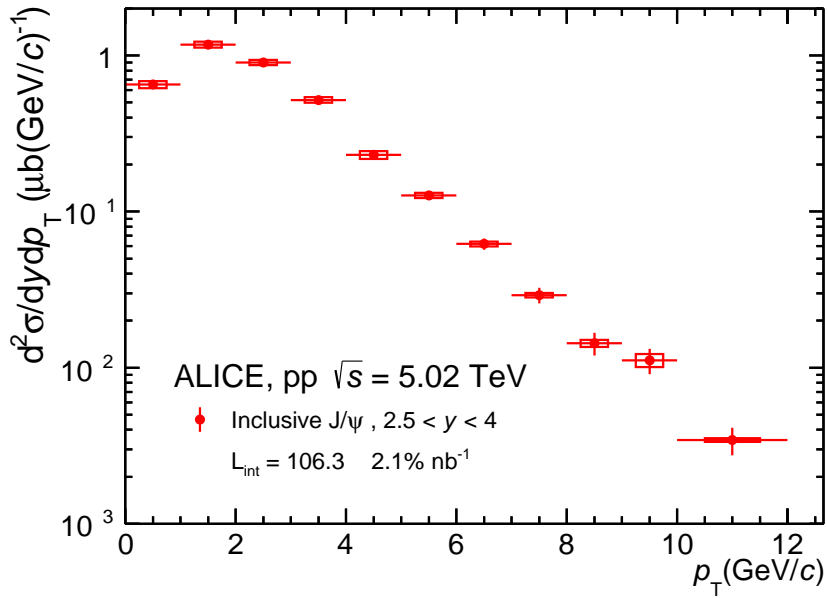} 
\caption{(colour online). The differential cross section ${\rm d}^2\sigma^{\rm pp}_{{\rm J}/\psi}/{\rm d}y{\rm d}p_{\rm T}$ for inclusive J/$\psi$ production in pp collisions at $\sqrt{s}=5.02$ TeV. The error bars represent the statistical uncertainties, the boxes around the points the uncorrelated systematic uncertainties. The uncertainty on the luminosity measurement represents a correlated global uncertainty.}
\label{fig:1bis}
\end{center}
\end{figure}

The $p_{\rm T}$-differential inclusive J/$\psi$ cross section in pp collisions at $\sqrt{s}=5.02$ TeV, in the region $2.5<y<4$, is shown in Fig.~\ref{fig:1bis}. The cross section value, integrated over the interval $2.5<y<4$, $p_{\rm T}<12$ GeV/$c$ is $\sigma^{\rm pp}_{{\rm J}/\psi} = 5.61 \pm 0.08 {\rm {(stat)}} \pm 0.28 {\rm {(syst)}}$ $\mu$b. These results are used as a reference in the determination of the nuclear modification factor for \pbpb\ collisions. Both the differential and integrated pp cross section values are consistent with those obtained via an interpolation~\cite{ALICE:2013spa,Adam:2015iga} of the measured values at 
$\sqrt{s}=2.76$ and 7 TeV~\cite{Abelev:2012kr,Abelev:2014qha}, which were used for the determination of the nuclear modification factor in p--Pb collisions at $\sqrt{s_{\rm NN}}=5.02$ TeV~\cite{Abelev:2013yxa,Adam:2015iga,Adam:2015jsa}.

The nuclear modification factor for inclusive J/$\psi$ production in \pbpb collisions at $\sqrt{s_{\rm NN}}=5.02$ TeV, integrated over the centrality range 0--90\%, and for the interval $2.5< y <4$, $p_{\rm T}<12$ GeV/$c$ is $R_{\rm AA}$($p_{\rm T}<12$ GeV/$c$) $ = 0.65\pm0.01 ({\rm stat}) \pm0.05 ({\rm syst})$, showing a significant suppression of the J/$\psi$ with respect to pp collisions at the same energy.
When restricting the $p_{\rm T}$ range to 8 GeV/$c$, corresponding to the interval covered in the $\sqrt{s_{\rm NN}}=2.76$ TeV results, one obtains $R_{\rm AA}$($p_{\rm T}<8$ GeV/$c$)$ = 0.66\pm0.01 ({\rm stat}) \pm0.05 ({\rm syst})$.
The ratio between the latter value and the corresponding one at $\snn=2.76$ TeV, $R_{\rm AA}$($p_{\rm T}<8$ GeV/$c$)$=0.58\pm0.01 ({\rm stat}) \pm0.09 ({\rm syst})$~\cite{Abelev:2013ila}, is $1.13\pm 0.02{\rm{(stat)}}\pm 0.18{\rm{(syst)}}$.
When calculating the ratio, the quoted uncertainties on the two values are considered as uncorrelated, except for the $\langle T_{\rm AA}\rangle$ contribution.

\begin{figure}
\begin{center}
\includegraphics[width=0.7\linewidth]{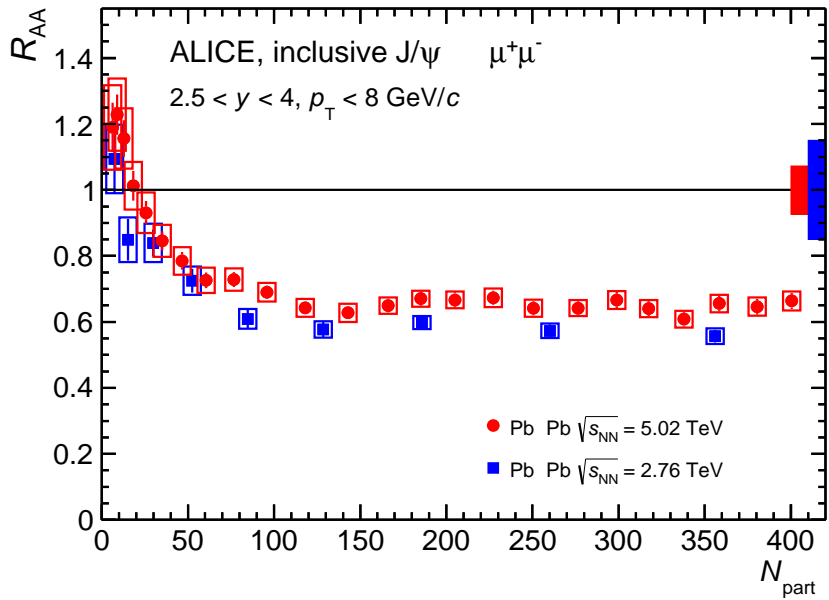} 
\caption{(colour online). The nuclear modification factor for inclusive J/$\psi$ production, as a function of centrality, at $\sqrt{s_{\rm NN}}=5.02$ TeV, compared to published results at $\sqrt{s_{\rm NN}}=2.76$ TeV~\cite{Abelev:2013ila}.
The error bars represent statistical uncertainties, the boxes around the points uncorrelated systematic uncertainties, while the centrality-correlated global uncertainties are shown as a filled box around $R_{\rm AA}=1$. The widths of the centrality classes used in the J/$\psi$ analysis at $\sqrt{s_{\rm NN}}=5.02$ TeV are 2\% from 0 to 12\%, then 3\% up to 30\% and 5\% for more peripheral collisions.}
\label{fig:2}
\end{center}
\end{figure}

Figure~\ref{fig:2} shows the centrality dependence of $R_{\rm AA}$ at $\sqrt{s_{\rm NN}}=5.02$ TeV. 
The results are compared to the values obtained at $\sqrt{s_{\rm NN}}=2.76$ TeV~\cite{Abelev:2013ila}, and correspond to the same  transverse-momentum range, $p_{\rm T}<8$ GeV/$c$. 
The centrality dependence, characterised by an increasing suppression with centrality up to $N_{\rm part}\sim 100$, followed by an approximately constant $R_{\rm AA}$ value, is similar at the two energies. 
A systematic difference by about 15\% is visible when comparing the two sets of results, even if the effect is within the total uncertainty of the measurements.  
The $R_{\rm AA}$ of prompt J/$\psi$ would be about 10\% higher if $R_{\rm AA}^{\text {non-prompt}}=0$ and about 5\% (1\%) smaller if $R_{\rm AA}^{\text  {non-prompt}}=1$ for central (peripheral) collisions.

\begin{figure}[h]
\begin{center}
\includegraphics[width=0.7\linewidth]{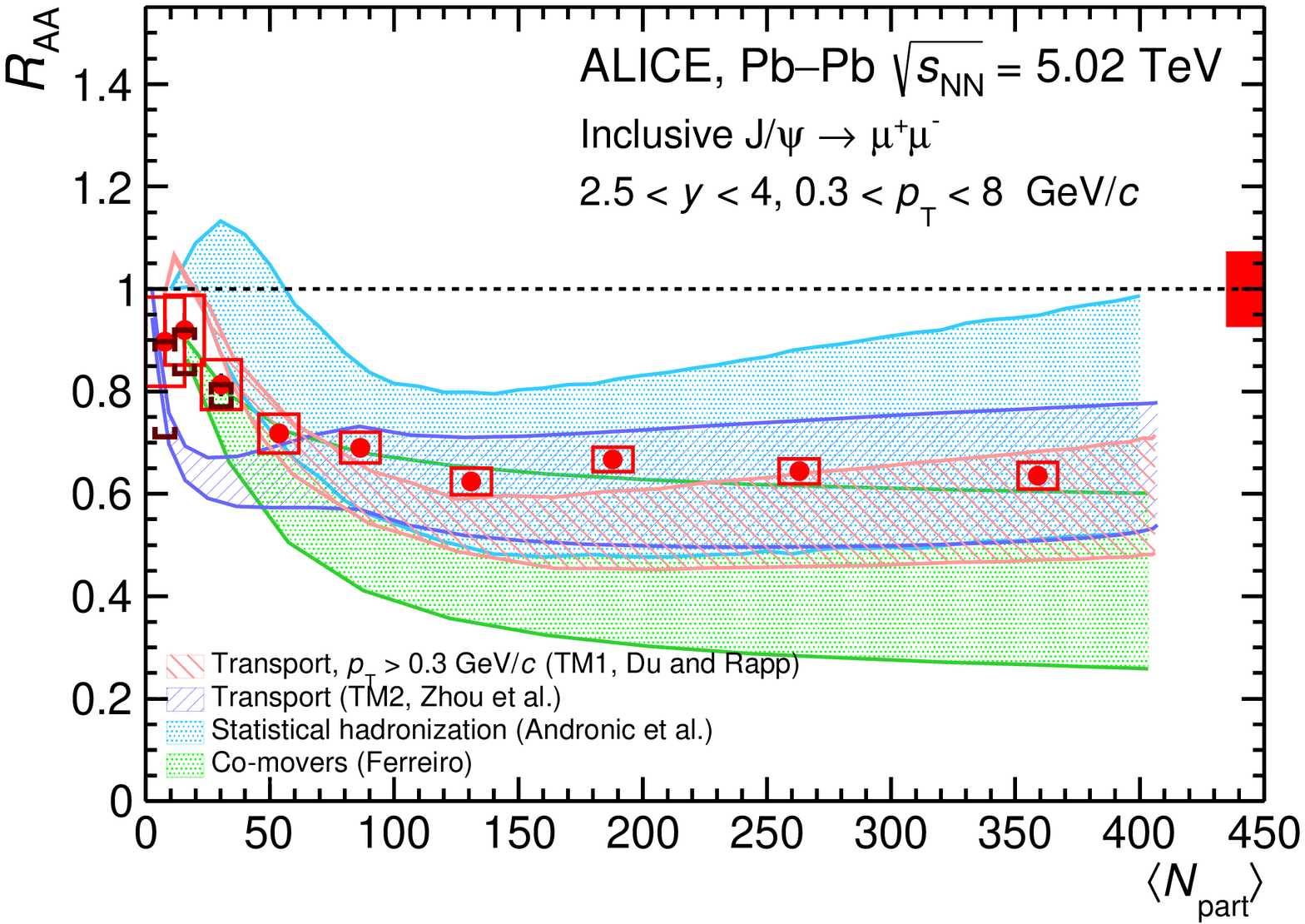} 
\caption{(colour online). Comparison of the centrality dependence (with 10\% width centrality classes) of the inclusive J/$\psi$ $R_{\rm AA}$ for $0.3<p_{\rm T}<8$ GeV/$c$ with theoretical models~\cite{Andronic:2012dm,Ferreiro:2012rq,Ferreiro:2014bia,Zhao:2011cv,Du:2015wha,Du:2016fam,Zhou:2014kka}. The model calculations do not include the $p_{\rm T}$ cut (except for TM1), which was anyway found to have a negligible impact, since they only include hadronic J/$\psi$ production. The error bars represent the statistical uncertainties, the boxes around the data points the uncorrelated systematic uncertainties, while the centrality-correlated global uncertainty is shown as a filled box around $R_{\rm AA}=1$. The brackets shown in the three most peripheral centrality
intervals represent the range of variation of the hadronic J/$\psi$ $R_{\rm AA}$ under extreme hypothesis on the photo-production contamination on the inclusive $R_{\rm AA}$.}
\label{fig:3}
\end{center}
\end{figure}

An excess of very-low $p_{\rm T}$ J/$\psi$, compared to the yield expected assuming a smooth evolution of the J/$\psi$ hadro-production and nuclear modification factor was observed in peripheral Pb--Pb collisions at $\sqrt{s_{\rm NN}}=2.76$ TeV~\cite{Adam:2015gba}. 
This excess might originate from the photo-production of J/$\psi$ and could influence the $R_{\rm AA}$ in peripheral collisions. 
To quantify the expected difference between the hadronic J/$\psi$ $R_{\rm AA}$ and the measured values the method described in~\cite{Adam:2015isa} was adopted. 
The hadronic J/$\psi$ $R_{\rm AA}$, for $0 < p_{\rm T} < 8$ GeV/$c$, is estimated to be about 34\%, 17\% and 9\% smaller than the measured values in the 80--90\%, 70--80\% and 60--70\% centrality classes, respectively. 
The variation decreases to about 9\%, 4\% and 2\%, respectively, when considering the $R_{\rm AA}$ for J/$\psi$ with $0.3 < p_{\rm T} < 8$ GeV/$c$, due to the remaining small contribution of photo-produced J/$\psi$.
Figure~\ref{fig:3} shows $R_{\rm AA}$ as a function of centrality, for $0.3<p_{\rm T}< 8$ GeV/$c$. 

Comparing the results of Fig.~\ref{fig:2} and Fig.~\ref{fig:3}, a less pronounced increase of $R_{\rm AA}$ for peripheral events can indeed be seen when such a selection is introduced.
The same extreme hypotheses as in~\cite{Adam:2015isa} were made to define upper and lower limits, represented with brackets on Fig.~\ref{fig:3}. 
Thus, the selection of J/$\psi$ with  $p_{\rm T}> 0.3$ GeV/$c$ makes the results more suitable for a comparison with theoretical models that only include hadronic J/$\psi$ production.

We start by comparing the results to a calculation based on a statistical model approach~\cite{Andronic:2012dm}, where J/$\psi$ are created, like all other hadrons, only at chemical freeze-out according to their statistical weights. 
In this model, the nucleon--nucleon ${\rm c{\overline c}}$ production cross section is extrapolated from LHCb pp measurements at $\sqrt{s}=7$ TeV~\cite{Aaij:2013mga} using FONLL calculations~\cite{Cacciari:2012ny}, obtaining ${\rm d}\sigma_{\rm c\overline c}/{\rm d}y = 0.45$ mb in the $y$ range covered by the data. 
Then, the nuclear modification of the parton distribution functions (shadowing) is accounted for via the EPS09 NLO parameterisation~\cite{Eskola:2009uj}.
The corresponding 17\% uncertainty on the extrapolated ${\rm d}\sigma_{\rm c\overline c}/{\rm d}y$ plus shadowing is used when calculating the uncertainty bands for this model. 
The results are also compared to the calculations of a transport model (TM1)~\cite{Zhao:2011cv,Du:2015wha,Du:2016fam} based on a thermal rate equation, which includes continuous dissociation and regeneration of the J/$\psi$ both in the QGP and in the hadronic phase. 
The inclusive ${\rm c{\overline c}}$ cross section is taken as ${\rm d}\sigma_{\rm c\overline c}/{\rm d}y=0.57$ mb, consistent with FONLL calculations, while the J/$\psi$ production cross section value in N--N collisions is ${\rm d}\sigma_{{\rm J/}\psi}/{\rm d}y=3.14$ $\mu$b.
The results of this model are shown as a band including a variation of the shadowing contribution between 10\% and 25\% and a 5\% uncertainty on the ${\rm c{\overline c}}$ cross section. 
The results are then compared to the calculations of a second transport model (TM2)~\cite{Zhou:2014kka}, which implements a hydrodynamic description of the medium evolution. The input nucleon--nucleon cross sections for $\rm c\overline c$ and J/$\psi$ are taken as ${\rm d}\sigma_{\rm c\overline c}/{\rm d}y=0.82$ mb, corresponding to the upper limit of FONLL calculations, and ${\rm d}\sigma_{{\rm J/}\psi}/{\rm d}y=3.5$ $\mu$b. Also for this model the band corresponds to the choice of either no shadowing, or a shadowing effect estimated with the EPS09 NLO parameterisation.
Finally, the data are compared to a `co-mover' model~\cite{Ferreiro:2012rq,Ferreiro:2014bia}, where the J/$\psi$ are dissociated via interactions with the partons/hadrons produced in the same rapidity range, using an effective interaction cross section $\sigma^{\textnormal{co-J}/\psi} = 0.65$ mb, based on 
calculations that described lower energy experimental results. Regeneration effects are included, based on ${\rm d}\sigma_{\rm c\overline c}/{\rm d}y$ values ranging from 0.45 to 0.7 mb, which correspond to the uncertainty band shown for the model. Shadowing effects, calculated within the Glauber-Gribov theory~\cite{Gribov:1968jf}, are included and are consistent with EKS98/nDSg predictions~\cite{Eskola:1998df,deFlorian:2003qf}. Finally, the contribution of non-prompt production is taken into account in the transport models TM1 and TM2, while it is not considered in the other calculations.

The data are described by the various calculations, the latter having rather large uncertainties, due to the choice of the corresponding input parameters, and in particular of ${\rm d}\sigma_{\rm c\overline c}/{\rm d}y$. It can be noted that for most calculations a better description is found when considering their upper limit. For transport models this corresponds to a minimum contribution or even absence of nuclear shadowing, which can be clearly considered as an extreme assumption for primary J/$\psi$, considering the J/$\psi$ measurements in p--Pb collisions~\cite{Adam:2015iga,Adam:2015jsa}.

\begin{figure}[h]
\begin{center}
\includegraphics[width=0.7\linewidth]{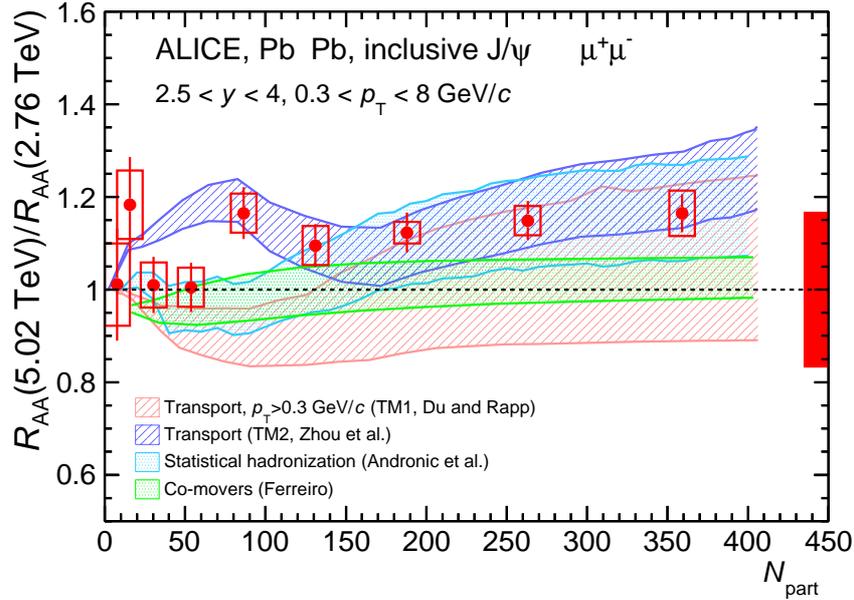} 
\caption{ (colour online). The ratio of the inclusive J/$\psi$ $R_{\rm AA}$ for $0.3<p_{\rm T}<8$ GeV/$c$ between $\sqrt{s_{\rm NN}}=5.02$ and 2.76 TeV, compared to theoretical models ~\cite{Andronic:2012dm,Ferreiro:2012rq,Ferreiro:2014bia,Zhao:2011cv,Du:2015wha,Du:2016fam,Zhou:2014kka}, shown as a function of centrality. The model calculations do not include the $p_{\rm T}$ cut (except for TM1), which was anyway found to have a negligible impact, since they only include  hadronic J/$\psi$ production. The error bars represent the statistical uncertainties and the boxes around the data points the uncorrelated systematic uncertainties. The centrality-correlated global uncertainty is shown as a filled box around $r=1$ and is obtained as the quadratic sum of the corresponding global uncertainties at $\sqrt{s_{\rm NN}} = 2.76$ and $5.02$ TeV.}
\label{fig:4}
\end{center}
\end{figure}

A correlation between the parameters of the models is present when comparing their calculations for $\snn = 2.76$ and  5.02 TeV. Therefore, the theoretical uncertainties can be reduced by forming the ratio $r=R_{\rm AA}({\rm 5.02\ TeV})/R_{\rm AA}({\rm 2.76\ TeV})$. Concerning data, the uncertainties on $\langle T_{\rm AA}\rangle$ cancel. In Fig.~\ref{fig:4} the centrality dependence of $r$, calculated for $0.3<p_{\rm T}<8$ GeV/$c$, is shown and compared to models. 
For prompt J/$\psi$ the ratio $r$ would be about 2\% (1--2\%) higher if beauty hadrons were fully (not) suppressed by the medium.
The transport model of Ref.~\cite{Zhao:2011cv,Du:2015wha,Du:2016fam} (TM1) shows a decrease of $r$ with increasing centrality, due to the larger suppression effects at high energy, followed by an increase, related to the effect of regeneration, which acts in the opposite direction and becomes dominant for central collisions. 
The other transport model (TM2)~\cite{Zhou:2014kka} also exhibits an increase for central collisions, while for peripheral collisions the behaviour is different. In the co-mover  model~\cite{Ferreiro:2012rq,Ferreiro:2014bia}, no structure is visible as a function of centrality, and the calculation favours $r$-values slightly below unity, implying that in this model the increase of the suppression effects with energy may be dominant over the regeneration effects for all centralities. 
Finally, the statistical model~\cite{Andronic:2012dm} shows a continuous increase of $r$ with centrality, dominated by the increase in the ${\rm c}\overline{\rm c}$ cross section with energy. 
The uncertainty bands shown in Fig.~\ref{fig:4} correspond to variations of about 5\% in the $\rm c\overline c$ cross section at $\sqrt{s_{\rm NN}}=5.02$ TeV, plus a 10\% relative variation of the shadowing contribution between the two energies in the case of TM1. 
The data are, within uncertainties, compatible with the theoretical models, and show no clear centrality dependence. The ratio for central collisions and $0.3<p_{\rm T}<8$ GeV/$c$ is  $r^{\rm 0-10\%} = 1.17 \pm 0.04 {\rm{(stat)}}\pm 0.20 {\rm{(syst)}}$. 

\begin{figure}
\begin{center}
\includegraphics[width=0.7\linewidth]{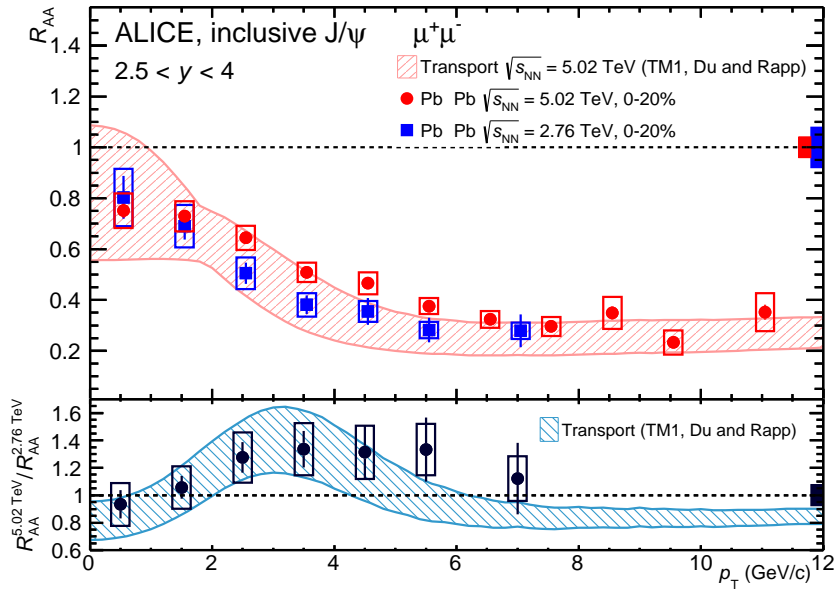} 
\caption{ (colour online). The $p_{\rm T}$ dependence of the inclusive J/$\psi$ $R_{\rm AA}$ at $\sqrt{s_{\rm NN}}=5.02$ TeV, compared to the corresponding result at $\sqrt{s_{\rm NN}}=2.76$ TeV~\cite{Abelev:2013ila} and to the calculation of a transport model~\cite{Zhao:2011cv,Du:2015wha,Du:2016fam}(TM1), in the centrality interval 0--20\%. The $p_{\rm T}$ dependence of $r$ is also shown for both data and theory. The error bars represent statistical uncertainties, the boxes around the points uncorrelated systematic uncertainties, while $p_{\rm T}$-correlated global uncertainties are shown as a filled box around $R_{\rm AA}=1$.}
\label{fig:5}
\end{center}
\end{figure}

Finally, the study of the $p_{\rm T}$ dependence of $R_{\rm AA}$ has proven to be a sensitive test of the presence of a regeneration component which, in calculations, leads to an increase at low $p_{\rm T}$. 
Figure~\ref{fig:5} shows, for the centrality interval 0--20\%, $R_{\rm AA}$ as a function of transverse momentum, compared to the corresponding results obtained at $\snn = 2.76$ TeV, and to a theoretical model calculation. 
The region $p_{\rm T}<0.3$ GeV/$c$ was not excluded, because the contribution of J/$\psi$ photo-production is negligible with respect to the hadronic one for central events~\cite{Adam:2015gba}. 
In the same figure the $p_{\rm T}$ dependence of $r$ is also shown.
A hint for an increase of $R_{\rm AA}$ with $\sqrt{s_{\rm NN}}$ is visible in the region $2<p_{\rm T}<6$ GeV/$c$, while the $r$-ratio is consistent with unity elsewhere.
This feature is qualitatively described by the theoretical model (TM1) also shown in the figure. 
The prompt J/$\psi$ $R_{\rm AA}$ is expected to be 7\% larger (2\% smaller) for $p_{\rm T} < 1$ GeV/$c$ and 30\% larger (55\% smaller) for $10 < p_{\rm T} < 12$ GeV/$c$ when the beauty contribution is fully (not) suppressed.
Assuming that $R_{\rm AA}^{\text {non-prompt}}$ does not vary significantly between the two collision energies, the ratio $r$ appears to be less sensitive to the non-prompt J/$\psi$ contribution. 
The effect is negligible for the case of full suppression of beauty hadrons, while it varies from no increase at low transverse momentum up to a maximum increase of about 15\% for $5 < p_{\rm T} < 6$ GeV/$c$ if no suppression is assumed.
The transport model of Ref.~\cite{Zhao:2011cv,Du:2015wha,Du:2016fam} (TM1) fairly describes the overall shape of the $R_{\rm AA}$ $p_{\rm T}$ dependence. 

\section{Conclusion}
We reported the ALICE measurement of inclusive \jpsi production in pp and \pbpb collisions at $\snn = 5.02$ TeV at the LHC. 
A systematic difference by about 15\% is visible when comparing the $R_{\rm AA}$ measured at $\snn = 5.02$ TeV to the one obtained at $\snn=2.76$ TeV, even if such an effect is within the total uncertainty of the measurements. 
When removing very-low $p_{\rm T}$ J/$\psi$ ($p_{\rm T}<0.3$ GeV/c), the $R_{\rm AA}$ shows a less pronounced increase for peripheral events, which can be ascribed to the removal of a large fraction of electromagnetic J/$\psi$ production~\cite{Adam:2015gba}. These results, as well as those on the ratio of the nuclear modification factors between $\sqrt{s_{\rm NN}}=5.02$ and 2.76 TeV, are described by theoretical calculations, and closer to their upper limits. 
The $p_{\rm T}$ dependence of $R_{\rm AA}$ exhibits an increase at low $p_{\rm T}$, a feature that in the model which is compared to the data is related to an important contribution of regenerated J/$\psi$. A hint for an increase of $R_{\rm AA}$ between $\sqrt{s_{\rm NN}}=2.76$ and 5.02 TeV is visible in the region $2<p_{\rm T}<6$ GeV/$c$, while the results are consistent elsewhere.
The results presented in this paper confirm that also at the highest energies reached today at the LHC, data on J/$\psi$ production support a picture
where a combination of suppression and regeneration takes place in the QGP, the two mechanisms being dominant at high and low $p_{\rm T}$, respectively.

%
%

\newenvironment{acknowledgement}{\relax}{\relax}
\begin{acknowledgement}
\section*{Acknowledgements}

The ALICE Collaboration would like to thank all its engineers and technicians for their invaluable contributions to the construction of the experiment and the CERN accelerator teams for the outstanding performance of the LHC complex.
The ALICE Collaboration gratefully acknowledges the resources and support provided by all Grid centres and the Worldwide LHC Computing Grid (WLCG) collaboration.
The ALICE Collaboration acknowledges the following funding agencies for their support in building and
running the ALICE detector:
State Committee of Science,  World Federation of Scientists (WFS)
and Swiss Fonds Kidagan, Armenia;
Conselho Nacional de Desenvolvimento Cient\'{\i}fico e Tecnol\'{o}gico (CNPq), Financiadora de Estudos e Projetos (FINEP),
Funda\c{c}\~{a}o de Amparo \`{a} Pesquisa do Estado de S\~{a}o Paulo (FAPESP);
Ministry of Science \& Technology of China (MSTC), National Natural Science Foundation of China (NSFC) and Ministry of Education of China (MOEC)";
Ministry of Science, Education and Sports of Croatia and  Unity through Knowledge Fund, Croatia;
Ministry of Education and Youth of the Czech Republic;
Danish Natural Science Research Council, the Carlsberg Foundation and the Danish National Research Foundation;
The European Research Council under the European Community's Seventh Framework Programme;
Helsinki Institute of Physics and the Academy of Finland;
French CNRS-IN2P3, the `Region Pays de Loire', `Region Alsace', `Region Auvergne' and CEA, France;
German Bundesministerium fur Bildung, Wissenschaft, Forschung und Technologie (BMBF) and the Helmholtz Association;
General Secretariat for Research and Technology, Ministry of Development, Greece;
National Research, Development and Innovation Office (NKFIH), Hungary;
Council of Scientific and Industrial Research (CSIR), New Delhi;
Department of Atomic Energy and Department of Science and Technology of the Government of India;
Istituto Nazionale di Fisica Nucleare (INFN) and Centro Fermi - Museo Storico della Fisica e Centro Studi e Ricerche ``Enrico Fermi'', Italy;
Japan Society for the Promotion of Science (JSPS) KAKENHI and MEXT, Japan;
National Research Foundation of Korea (NRF);
Consejo Nacional de Cienca y Tecnologia (CONACYT), Direccion General de Asuntos del Personal Academico(DGAPA), M\'{e}xico, Amerique Latine Formation academique - 
European Commission~(ALFA-EC) and the EPLANET Program~(European Particle Physics Latin American Network);
Stichting voor Fundamenteel Onderzoek der Materie (FOM) and the Nederlandse Organisatie voor Wetenschappelijk Onderzoek (NWO), Netherlands;
Research Council of Norway (NFR);
Pontificia Universidad Cat\'{o}lica del Per\'{u};
National Science Centre, Poland;
Ministry of National Education/Institute for Atomic Physics and National Council of Scientific Research in Higher Education~(CNCSI-UEFISCDI), Romania;
Joint Institute for Nuclear Research, Dubna;
Ministry of Education and Science of Russian Federation, Russian Academy of Sciences, Russian Federal Agency of Atomic Energy, Russian Federal Agency for Science and Innovations and The Russian Foundation for Basic Research;
Ministry of Education of Slovakia;
Department of Science and Technology, South Africa;
Centro de Investigaciones Energeticas, Medioambientales y Tecnologicas (CIEMAT), E-Infrastructure shared between Europe and Latin America (EELA), 
Ministerio de Econom\'{i}a y Competitividad (MINECO) of Spain, Xunta de Galicia (Conseller\'{\i}a de Educaci\'{o}n),
Centro de Aplicaciones Tecnológicas y Desarrollo Nuclear (CEA\-DEN), Cubaenerg\'{\i}a, Cuba, and IAEA (International Atomic Energy Agency);
Swedish Research Council (VR) and Knut $\&$ Alice Wallenberg Foundation (KAW);
National Science and Technology Development Agency (NSDTA), Suranaree University of Technology (SUT) and Office of the Higher Education Commission under NRU project of Thailand;
Ukraine Ministry of Education and Science;
United Kingdom Science and Technology Facilities Council (STFC);
The United States Department of Energy, the United States National Science Foundation, the State of Texas, and the State of Ohio.

\end{acknowledgement}

\bibliographystyle{utphys}   
\bibliography{JpsiPbPb15_CernPreprint}

\providecommand{\href}[2]{#2}\begingroup\raggedright\begin{thebibliography}{10}

\bibitem{Shuryak:1978ij}
E.~V. Shuryak, ``{Quark-Gluon Plasma and Hadronic Production of Leptons,
  Photons and Psions},''
  \href{http://dx.doi.org/10.1016/0370-2693(78)90370-2}{{\em Phys. Lett.}
  {\bfseries B78} (1978) 150}.
[Yad. Fiz.28,796(1978)].

\bibitem{Heinz:2000bk}
U.~W. Heinz and M.~Jacob, ``{Evidence for a new state of matter: An assessment
  of the results from the CERN lead beam program},''
\href{http://arxiv.org/abs/nucl-th/0002042}{{\ttfamily arXiv:nucl-th/0002042
  [nucl-th]}}.

\bibitem{Arsene:2004fa}
{\bfseries BRAHMS} Collaboration, I.~Arsene {\em et~al.}, ``{Quark gluon plasma
  and color glass condensate at RHIC? The Perspective from the BRAHMS
  experiment},'' \href{http://dx.doi.org/10.1016/j.nuclphysa.2005.02.130}{{\em
  Nucl. Phys.} {\bfseries A757} (2005) 1--27},
\href{http://arxiv.org/abs/nucl-ex/0410020}{{\ttfamily arXiv:nucl-ex/0410020
  [nucl-ex]}}.

\bibitem{Back:2004je}
{\bfseries PHOBOS} Collaboration, B.~B. Back {\em et~al.}, ``{The PHOBOS
  perspective on discoveries at RHIC},''
  \href{http://dx.doi.org/10.1016/j.nuclphysa.2005.03.084}{{\em Nucl. Phys.}
  {\bfseries A757} (2005) 28--101},
\href{http://arxiv.org/abs/nucl-ex/0410022}{{\ttfamily arXiv:nucl-ex/0410022
  [nucl-ex]}}.

\bibitem{Adams:2005dq}
{\bfseries STAR} Collaboration, J.~Adams {\em et~al.}, ``{Experimental and
  theoretical challenges in the search for the quark gluon plasma: The STAR
  Collaboration's critical assessment of the evidence from RHIC collisions},''
  \href{http://dx.doi.org/10.1016/j.nuclphysa.2005.03.085}{{\em Nucl. Phys.}
  {\bfseries A757} (2005) 102--183},
\href{http://arxiv.org/abs/nucl-ex/0501009}{{\ttfamily arXiv:nucl-ex/0501009
  [nucl-ex]}}.

\bibitem{Adcox:2004mh}
{\bfseries PHENIX} Collaboration, K.~Adcox {\em et~al.}, ``{Formation of dense
  partonic matter in relativistic nucleus-nucleus collisions at RHIC:
  Experimental evaluation by the PHENIX collaboration},''
  \href{http://dx.doi.org/10.1016/j.nuclphysa.2005.03.086}{{\em Nucl. Phys.}
  {\bfseries A757} (2005) 184--283},
\href{http://arxiv.org/abs/nucl-ex/0410003}{{\ttfamily arXiv:nucl-ex/0410003
  [nucl-ex]}}.

\bibitem{Muller:2012zq}
B.~M\"uller, J.~Schukraft, and B.~Wyslouch, ``{First Results from Pb+Pb
  collisions at the LHC},''
  \href{http://dx.doi.org/10.1146/annurev-nucl-102711-094910}{{\em Ann. Rev.
  Nucl. Part. Sci.} {\bfseries 62} (2012) 361--386},
\href{http://arxiv.org/abs/1202.3233}{{\ttfamily arXiv:1202.3233 [hep-ex]}}.

\bibitem{Brambilla:2010cs}
N.~Brambilla, S.~Eidelman, B.~Heltsley, R.~Vogt, G.~Bodwin, {\em et~al.},
  ``{Heavy quarkonium: progress, puzzles, and opportunities},''
  \href{http://dx.doi.org/10.1140/epjc/s10052-010-1534-9}{{\em Eur. Phys. J.}
  {\bfseries C71} (2011) 1534},
\href{http://arxiv.org/abs/1010.5827}{{\ttfamily arXiv:1010.5827 [hep-ph]}}.

\bibitem{Alessandro:2004ap}
{\bfseries NA50} Collaboration, B.~Alessandro {\em et~al.}, ``{A new
  measurement of J/$\psi$ suppression in Pb-Pb collisions at 158-GeV per
  nucleon},'' \href{http://dx.doi.org/10.1140/epjc/s2004-02107-9}{{\em Eur.
  Phys. J.} {\bfseries C39} (2005) 335--345},
\href{http://arxiv.org/abs/hep-ex/0412036}{{\ttfamily arXiv:hep-ex/0412036
  [hep-ex]}}.

\bibitem{Arnaldi:2007zz}
{\bfseries NA60} Collaboration, R.~Arnaldi {\em et~al.}, ``{J/$\psi$ production
  in indium-indium collisions at 158-GeV/nucleon},''
\href{http://dx.doi.org/10.1103/PhysRevLett.99.132302}{{\em Phys. Rev. Lett.}
  {\bfseries 99} (2007) 132302}.

\bibitem{Adare:2011yf}
{\bfseries PHENIX} Collaboration, A.~Adare {\em et~al.}, ``{$J/\psi$
  suppression at forward rapidity in Au+Au collisions at $\sqrt{s_{NN}}=200$
  GeV},'' \href{http://dx.doi.org/10.1103/PhysRevC.84.054912}{{\em Phys. Rev.}
  {\bfseries C84} (2011) 054912},
\href{http://arxiv.org/abs/1103.6269}{{\ttfamily arXiv:1103.6269 [nucl-ex]}}.

\bibitem{Abelev:2009qaa}
{\bfseries STAR} Collaboration, B.~I. Abelev {\em et~al.}, ``{J/$\psi$
  production at high transverse momentum in p+p and Cu+Cu collisions at
  $\sqrt{s_{\rm NN}} = 200$ GeV},''
  \href{http://dx.doi.org/10.1103/PhysRevC.80.041902}{{\em Phys. Rev.}
  {\bfseries C80} (2009) 041902},
\href{http://arxiv.org/abs/0904.0439}{{\ttfamily arXiv:0904.0439 [nucl-ex]}}.

\bibitem{Abelev:2012rv}
{\bfseries ALICE} Collaboration, B.~Abelev {\em et~al.}, ``{J/$\psi$
  Suppression at Forward Rapidity in Pb-Pb Collisions at $\sqrt{s_{{\rm NN}}}
  =2.76$ TeV},'' \href{http://dx.doi.org/10.1103/PhysRevLett.109.072301}{{\em
  Phys. Rev. Lett.} {\bfseries 109} no.~7, (Aug., 2012) 072301},
  \href{http://arxiv.org/abs/1202.1383}{{\ttfamily arXiv:1202.1383 [hep-exp]}}.

\bibitem{Chatrchyan:2012np}
{\bfseries CMS} Collaboration, S.~Chatrchyan {\em et~al.}, ``{Suppression of
  non-prompt $J/\psi$, prompt $J/\psi$, and Y(1S) in PbPb collisions at
  $\sqrt{s_{NN}}=2.76$ TeV},''
  \href{http://dx.doi.org/10.1007/JHEP05(2012)063}{{\em JHEP} {\bfseries 05}
  (2012) 063},
\href{http://arxiv.org/abs/1201.5069}{{\ttfamily arXiv:1201.5069 [nucl-ex]}}.

\bibitem{Matsui:1986dk}
T.~Matsui and H.~Satz, ``{$J/\psi$ Suppression by Quark-Gluon Plasma
  Formation},'' \href{http://dx.doi.org/10.1016/0370-2693(86)91404-8}{{\em
  Phys. Lett.} {\bfseries B178} (1986) 416}.

\bibitem{Digal:2001ue}
S.~Digal, P.~Petreczky, and H.~Satz, ``{Quarkonium feed down and sequential
  suppression},'' \href{http://dx.doi.org/10.1103/PhysRevD.64.094015}{{\em
  Phys. Rev.} {\bfseries D64} (2001) 094015},
\href{http://arxiv.org/abs/hep-ph/0106017}{{\ttfamily arXiv:hep-ph/0106017
  [hep-ph]}}.

\bibitem{Ferreiro:2012rq}
E.~G. Ferreiro, ``{Charmonium dissociation and recombination at LHC: Revisiting
  comovers},'' \href{http://dx.doi.org/10.1016/j.physletb.2014.02.011}{{\em
  Phys. Lett.} {\bfseries B731} (2014) 57--63},
\href{http://arxiv.org/abs/1210.3209}{{\ttfamily arXiv:1210.3209 [hep-ph]}}.

\bibitem{Zhao:2011cv}
X.~Zhao and R.~Rapp, ``{Medium Modifications and Production of Charmonia at
  LHC},'' \href{http://dx.doi.org/10.1016/j.nuclphysa.2011.05.001}{{\em Nucl.
  Phys.} {\bfseries A859} (2011) 114--125},
\href{http://arxiv.org/abs/1102.2194}{{\ttfamily arXiv:1102.2194 [hep-ph]}}.

\bibitem{Zhou:2014kka}
K.~Zhou, N.~Xu, Z.~Xu, and P.~Zhuang, ``{Medium effects on charmonium
  production at ultrarelativistic energies available at the CERN Large Hadron
  Collider},'' \href{http://dx.doi.org/10.1103/PhysRevC.89.054911}{{\em Phys.
  Rev.} {\bfseries C89} no.~5, (2014) 054911},
\href{http://arxiv.org/abs/1401.5845}{{\ttfamily arXiv:1401.5845 [nucl-th]}}.

\bibitem{Abelev:2013ila}
{\bfseries ALICE} Collaboration, B.~Abelev {\em et~al.}, ``{Centrality,
  rapidity and transverse momentum dependence of $J/\Psi$ suppression in Pb-Pb
  collisions at $\sqrt{s_{NN}}$=2.76 TeV},''
  \href{http://dx.doi.org/10.1016/j.physletb.2014.05.064}{{\em Phys. Lett.}
  {\bfseries B743} (2014) 314--327},
\href{http://arxiv.org/abs/1311.0214}{{\ttfamily arXiv:1311.0214 [nucl-ex]}}.

\bibitem{Adam:2015isa}
{\bfseries ALICE} Collaboration, J.~Adam {\em et~al.}, ``{Differential studies
  of inclusive J/$\psi$ and $\psi$(2S) production at forward rapidity in Pb-Pb
  collisions at $ \sqrt{s_{\mathrm{NN}}}=2.76 $ TeV},''
  \href{http://dx.doi.org/10.1007/JHEP05(2016)179}{{\em JHEP} {\bfseries 05}
  (2016) 179},
\href{http://arxiv.org/abs/1506.08804}{{\ttfamily arXiv:1506.08804 [nucl-ex]}}.

\bibitem{Adam:2015lda}
{\bfseries ALICE} Collaboration, J.~Adam {\em et~al.}, ``{Direct photon
  production in Pb-Pb collisions at $\sqrt{s_\mathrm{NN}} =$ 2.76 TeV},''
  \href{http://dx.doi.org/10.1016/j.physletb.2016.01.020}{{\em Phys. Lett.}
  {\bfseries B754} (2016) 235--248},
\href{http://arxiv.org/abs/1509.07324}{{\ttfamily arXiv:1509.07324 [nucl-ex]}}.

\bibitem{BraunMunzinger:2000px}
P.~Braun-Munzinger and J.~Stachel, ``{(Non)thermal aspects of charmonium
  production and a new look at $J/\psi$ suppression},''
\href{http://dx.doi.org/10.1016/S0370-2693(00)00991-6}{{\em Phys. Lett.}
  {\bfseries B490} (2000) 196--202}.

\bibitem{Thews:2000rj}
R.~L. Thews, M.~Schroedter, and J.~Rafelski, ``{Enhanced $J/\psi$ production in
  deconfined quark matter},''
  \href{http://dx.doi.org/10.1103/PhysRevC.63.054905}{{\em Phys. Rev.}
  {\bfseries C63} (2001) 054905},
\href{http://arxiv.org/abs/hep-ph/0007323}{{\ttfamily arXiv:hep-ph/0007323
  [hep-ph]}}.

\bibitem{Bhattacharya:2014ara}
T.~Bhattacharya {\em et~al.}, ``{QCD Phase Transition with Chiral Quarks and
  Physical Quark Masses},''
  \href{http://dx.doi.org/10.1103/PhysRevLett.113.082001}{{\em Phys. Rev.
  Lett.} {\bfseries 113} no.~8, (2014) 082001},
\href{http://arxiv.org/abs/1402.5175}{{\ttfamily arXiv:1402.5175 [hep-lat]}}.

\bibitem{Aamodt:2008zz}
{\bfseries ALICE} Collaboration, K.~Aamodt {\em et~al.}, ``{The ALICE
  experiment at the CERN LHC},''
\href{http://dx.doi.org/10.1088/1748-0221/3/08/S08002}{{\em JINST} {\bfseries
  3} (2008) S08002}.

\bibitem{Abelev:2014ffa}
{\bfseries ALICE} Collaboration, B.~Abelev {\em et~al.}, ``{Performance of the
  ALICE Experiment at the CERN LHC},'' {\em Int. J. Mod. Phys.} {\bfseries A29}
  (2014) 1430044,
\href{http://arxiv.org/abs/1402.4476}{{\ttfamily arXiv:1402.4476 [nucl-ex]}}.

\bibitem{Aamodt:2011gj}
{\bfseries ALICE} Collaboration, K.~Aamodt {\em et~al.}, ``{Rapidity and
  transverse momentum dependence of inclusive J/$\psi$ production in pp
  collisions at $\sqrt{s} = 7$ TeV},''
  \href{http://dx.doi.org/10.1016/j.physletb.2011.09.054,
  10.1016/j.physletb.2012.10.060}{{\em Phys. Lett.} {\bfseries B704} (2011)
  442--455},
\href{http://arxiv.org/abs/1105.0380}{{\ttfamily arXiv:1105.0380 [hep-ex]}}.

\bibitem{Aamodt:2010aa}
{\bfseries ALICE} Collaboration, K.~Aamodt {\em et~al.}, ``{Alignment of the
  ALICE Inner Tracking System with cosmic-ray tracks},''
  \href{http://dx.doi.org/10.1088/1748-0221/5/03/P03003}{{\em JINST} {\bfseries
  5} (2010) P03003},
\href{http://arxiv.org/abs/1001.0502}{{\ttfamily arXiv:1001.0502
  [physics.ins-det]}}.

\bibitem{Abbas:2013taa}
{\bfseries ALICE} Collaboration, E.~Abbas {\em et~al.}, ``{Performance of the
  ALICE VZERO system},''
  \href{http://dx.doi.org/10.1088/1748-0221/8/10/P10016}{{\em JINST} {\bfseries
  8} (2013) P10016},
\href{http://arxiv.org/abs/1306.3130}{{\ttfamily arXiv:1306.3130 [nucl-ex]}}.

\bibitem{Bondila:2005xy}
M.~Bondila {\em et~al.}, ``{ALICE T0 detector},''
\href{http://dx.doi.org/10.1109/TNS.2005.856900}{{\em IEEE Trans. Nucl. Sci.}
  {\bfseries 52} (2005) 1705--1711}.

\bibitem{ALICE:2012aa}
{\bfseries ALICE} Collaboration, B.~Abelev {\em et~al.}, ``{Measurement of the
  Cross Section for Electromagnetic Dissociation with Neutron Emission in Pb-Pb
  Collisions at $\sqrt{s_{NN}}$ = 2.76 TeV},''
  \href{http://dx.doi.org/10.1103/PhysRevLett.109.252302}{{\em Phys. Rev.
  Lett.} {\bfseries 109} (2012) 252302},
\href{http://arxiv.org/abs/1203.2436}{{\ttfamily arXiv:1203.2436 [nucl-ex]}}.

\bibitem{Abelev:2013qoq}
{\bfseries ALICE} Collaboration, B.~Abelev {\em et~al.}, ``{Centrality
  determination of Pb-Pb collisions at $\sqrt{s_{NN}}$ = 2.76 TeV with
  ALICE},'' \href{http://dx.doi.org/10.1103/PhysRevC.88.044909}{{\em Phys.
  Rev.} {\bfseries C88} no.~4, (2013) 044909},
\href{http://arxiv.org/abs/1301.4361}{{\ttfamily arXiv:1301.4361 [nucl-ex]}}.

\bibitem{Adam:2015ptt}
{\bfseries ALICE} Collaboration, J.~Adam {\em et~al.}, ``{Centrality dependence
  of the charged-particle multiplicity density at midrapidity in Pb-Pb
  collisions at $\sqrt{s_{\rm NN}}$ = 5.02 TeV},''
  \href{http://dx.doi.org/10.1103/PhysRevLett.116.222302}{{\em Phys. Rev.
  Lett.} {\bfseries 116} no.~22, (2016) 222302},
\href{http://arxiv.org/abs/1512.06104}{{\ttfamily arXiv:1512.06104 [nucl-ex]}}.

\bibitem{ALICE-PUBLIC-2015-006}
{ALICE Collaboration}, ``{Quarkonium signal extraction in ALICE},'' {\em
  {\href{https://cds.cern.ch/record/2060096}{ALICE-PUBLIC-2015-006}}} (2015) .

\bibitem{Abelev:2014zpa}
{\bfseries ALICE} Collaboration, B.~Abelev {\em et~al.}, ``{Suppression of
  $\psi$(2S) production in p-Pb collisions at $\sqrt{s_{\rm NN}}$ = 5.02
  TeV},'' \href{http://dx.doi.org/10.1007/JHEP12(2014)073}{{\em JHEP}
  {\bfseries 12} (2014) 073},
\href{http://arxiv.org/abs/1405.3796}{{\ttfamily arXiv:1405.3796 [nucl-ex]}}.

\bibitem{Agostinelli:2002hh}
{\bfseries GEANT4} Collaboration, S.~Agostinelli {\em et~al.}, ``{GEANT4: A
  Simulation toolkit},''
\href{http://dx.doi.org/10.1016/S0168-9002(03)01368-8}{{\em Nucl. Instrum.
  Meth.} {\bfseries A506} (2003) 250--303}.

\bibitem{Brun:1994aa}
R.~Brun, F.~Carminati, and S.~Giani, ``{GEANT Detector Description and
  Simulation Tool},''
{\em CERN Program Library Long Writeup} {\bfseries CERN-W5013} (1994) .

\bibitem{Agashe:2014kda}
{\bfseries Particle Data Group} Collaboration, K.~A. Olive {\em et~al.},
  ``{Review of Particle Physics},''
\href{http://dx.doi.org/10.1088/1674-1137/38/9/090001}{{\em Chin. Phys.}
  {\bfseries C38} (2014) 090001}.

\bibitem{Abelev:2013yxa}
{\bfseries ALICE} Collaboration, B.~Abelev {\em et~al.}, ``{J/$\psi$ production
  and nuclear effects in p-Pb collisions at $\sqrt{s_{{\rm NN}}} = 5.02$
  TeV},'' \href{http://dx.doi.org/10.1007/JHEP02(2014)073}{{\em JHEP}
  {\bfseries 1402} (2014) 073},
\href{http://arxiv.org/abs/1308.6726}{{\ttfamily arXiv:1308.6726 [nucl-ex]}}.

\bibitem{ALICE-PUBLIC-2015-008}
{\bfseries ALICE} Collaboration, ``{Centrality dependence of the
  charged-particle multiplicity density at midrapidity in Pb-Pb collisions at
  $\sqrt{s_{\rm NN}}$ = 5.02 TeV},''. \url{https://cds.cern.ch/record/2118084}.

\bibitem{Abelev:2011md}
{\bfseries ALICE} Collaboration, B.~Abelev {\em et~al.}, ``{J/$\psi$
  polarization in pp collisions at $\sqrt{s}=7$ TeV},''
  \href{http://dx.doi.org/10.1103/PhysRevLett.108.082001}{{\em Phys. Rev.
  Lett.} {\bfseries 108} (2012) 082001},
\href{http://arxiv.org/abs/1111.1630}{{\ttfamily arXiv:1111.1630 [hep-ex]}}.

\bibitem{Aaij:2011jh}
{\bfseries LHCb} Collaboration, R.~Aaij {\em et~al.}, ``{Measurement of
  $J/\psi$ production in $pp$ collisions at $\sqrt{s}=7~\rm{TeV}$},''
  \href{http://dx.doi.org/10.1140/epjc/s10052-011-1645-y}{{\em Eur. Phys. J.}
  {\bfseries C71} (2011) 1645},
\href{http://arxiv.org/abs/1103.0423}{{\ttfamily arXiv:1103.0423 [hep-ex]}}.

\bibitem{ALICE-PUBLIC-2016-005}
{\bfseries ALICE} Collaboration, ``{ALICE luminosity determination for pp
  collisions at $\sqrt{s}=5$ TeV},''. \url{https://cds.cern.ch/record/2202638}.

\bibitem{ALICE:2013spa}
{{\bf ALICE} and {\bf LHCb} Collaborations}, ``{Reference $pp$ cross-sections
  for $J/\psi$ studies in proton-lead collisions at $\sqrt{s_{NN}} = 5.02$~TeV
  and comparisons between ALICE and LHCb results},''
{\em {\href{https://cds.cern.ch/record/1639617}{ALICE-PUBLIC-2013-002},
  LHCb-CONF-2013-013}} (2013) .

\bibitem{Aaij:2012asz}
{\bfseries LHCb} Collaboration, R.~Aaij {\em et~al.}, ``{Measurement of
  $J/\psi$ production in $pp$ collisions at $\sqrt{s}=2.76$ TeV},''
  \href{http://dx.doi.org/10.1007/JHEP02(2013)041}{{\em JHEP} {\bfseries 02}
  (2013) 041},
\href{http://arxiv.org/abs/1212.1045}{{\ttfamily arXiv:1212.1045 [hep-ex]}}.

\bibitem{Adam:2015iga}
{\bfseries ALICE} Collaboration, J.~Adam {\em et~al.}, ``{Rapidity and
  transverse-momentum dependence of the inclusive J/$\psi$ nuclear modification
  factor in p-Pb collisions at $ \sqrt{s_{N\ N}} =$ 5.02 TeV},''
  \href{http://dx.doi.org/10.1007/JHEP06(2015)055}{{\em JHEP} {\bfseries 06}
  (2015) 055},
\href{http://arxiv.org/abs/1503.07179}{{\ttfamily arXiv:1503.07179 [nucl-ex]}}.

\bibitem{Abelev:2012kr}
{\bfseries ALICE} Collaboration, B.~Abelev {\em et~al.}, ``{Inclusive J/$\psi$
  production in $pp$ collisions at $\sqrt{s} = 2.76$ TeV},''
  \href{http://dx.doi.org/10.1016/j.physletb.2012.10.078,
  10.1016/j.physletb.2015.06.058}{{\em Phys. Lett.} {\bfseries B718} (2012)
  295--306},
\href{http://arxiv.org/abs/1203.3641}{{\ttfamily arXiv:1203.3641 [hep-ex]}}.

\bibitem{Abelev:2014qha}
{\bfseries ALICE} Collaboration, B.~Abelev {\em et~al.}, ``{Measurement of
  quarkonium production at forward rapidity in pp collisions at $\sqrt{s}$= 7
  TeV},'' \href{http://dx.doi.org/10.1140/epjc/s10052-014-2974-4}{{\em Eur.
  Phys. J.} {\bfseries C74} (2014) 2974},
\href{http://arxiv.org/abs/1403.3648}{{\ttfamily arXiv:1403.3648 [nucl-ex]}}.

\bibitem{Adam:2015jsa}
{\bfseries ALICE} Collaboration, J.~Adam {\em et~al.}, ``{Centrality dependence
  of inclusive J/$\psi$ production in p-Pb collisions at $
  \sqrt{s_{\mathrm{NN}}}=5.02 $ TeV},''
  \href{http://dx.doi.org/10.1007/JHEP11(2015)127}{{\em JHEP} {\bfseries 11}
  (2015) 127},
\href{http://arxiv.org/abs/1506.08808}{{\ttfamily arXiv:1506.08808 [nucl-ex]}}.

\bibitem{Adam:2015gba}
{\bfseries ALICE} Collaboration, J.~Adam {\em et~al.}, ``{Measurement of an
  excess in the yield of J/$\psi$ at very low $p_{\rm T}$ in Pb-Pb collisions
  at $\sqrt{s_{\rm NN}}$ = 2.76 TeV},''
  \href{http://dx.doi.org/10.1103/PhysRevLett.116.222301}{{\em Phys. Rev.
  Lett.} {\bfseries 116} (2016) 222301},
\href{http://arxiv.org/abs/1509.08802}{{\ttfamily arXiv:1509.08802 [nucl-ex]}}.

\bibitem{Andronic:2012dm}
A.~Andronic, P.~Braun-Munzinger, K.~Redlich, and J.~Stachel, ``{The statistical
  model in Pb-Pb collisions at the LHC},''
  \href{http://dx.doi.org/10.1016/j.nuclphysa.2013.02.070}{{\em Nucl. Phys.}
  {\bfseries A904-905} (2013) 535c--538c},
\href{http://arxiv.org/abs/1210.7724}{{\ttfamily arXiv:1210.7724 [nucl-th]}}.

\bibitem{Ferreiro:2014bia}
E.~G. Ferreiro, ``{Excited charmonium suppression in proton-nucleus collisions
  as a consequence of comovers},''
  \href{http://dx.doi.org/10.1016/j.physletb.2015.07.066}{{\em Phys. Lett.}
  {\bfseries B749} (2015) 98--103},
\href{http://arxiv.org/abs/1411.0549}{{\ttfamily arXiv:1411.0549 [hep-ph]}}.

\bibitem{Du:2015wha}
X.~Du and R.~Rapp, ``{Sequential Regeneration of Charmonia in Heavy-Ion
  Collisions},'' \href{http://dx.doi.org/10.1016/j.nuclphysa.2015.09.006}{{\em
  Nucl. Phys.} {\bfseries A943} (2015) 147--158},
\href{http://arxiv.org/abs/1504.00670}{{\ttfamily arXiv:1504.00670 [hep-ph]}}.

\bibitem{Du:2016fam}
X.~Du and R.~Rapp, ``{$\psi$(2S) Production at the LHC},'' in {\em {16th
  International Conference on Strangeness in Quark Matter (SQM 2016) Berkeley,
  California, United States, June 27-July 1, 2016}}.
\newblock 2016.
\newblock
\href{http://arxiv.org/abs/1609.04868}{{\ttfamily arXiv:1609.04868 [hep-ph]}}.
\newblock

\bibitem{Aaij:2013mga}
{\bfseries LHCb} Collaboration, R.~Aaij {\em et~al.}, ``{Prompt charm
  production in pp collisions at $\sqrt{s}=7$ TeV},''
  \href{http://dx.doi.org/10.1016/j.nuclphysb.2013.02.010}{{\em Nucl. Phys.}
  {\bfseries B871} (2013) 1--20},
\href{http://arxiv.org/abs/1302.2864}{{\ttfamily arXiv:1302.2864 [hep-ex]}}.

\bibitem{Cacciari:2012ny}
M.~Cacciari, S.~Frixione, N.~Houdeau, M.~L. Mangano, P.~Nason, and G.~Ridolfi,
  ``{Theoretical predictions for charm and bottom production at the LHC},''
  \href{http://dx.doi.org/10.1007/JHEP10(2012)137}{{\em JHEP} {\bfseries 10}
  (2012) 137},
\href{http://arxiv.org/abs/1205.6344}{{\ttfamily arXiv:1205.6344 [hep-ph]}}.

\bibitem{Eskola:2009uj}
K.~J. Eskola, H.~Paukkunen, and C.~A. Salgado, ``{EPS09: A New Generation of
  NLO and LO Nuclear Parton Distribution Functions},''
  \href{http://dx.doi.org/10.1088/1126-6708/2009/04/065}{{\em JHEP} {\bfseries
  0904} (2009) 065},
\href{http://arxiv.org/abs/0902.4154}{{\ttfamily arXiv:0902.4154 [hep-ph]}}.

\bibitem{Gribov:1968jf}
V.~N. Gribov, ``{Glauber corrections and the interaction between high-energy
  hadrons and nuclei},'' {\em Sov. Phys. JETP} {\bfseries 29} (1969) 483--487.
[Zh. Eksp. Teor. Fiz.56,892(1969)].

\bibitem{Eskola:1998df}
K.~J. Eskola, V.~J. Kolhinen, and C.~A. Salgado, ``{The Scale dependent nuclear
  effects in parton distributions for practical applications},''
  \href{http://dx.doi.org/10.1007/s100520050513}{{\em Eur. Phys. J.} {\bfseries
  C9} (1999) 61--68},
\href{http://arxiv.org/abs/hep-ph/9807297}{{\ttfamily arXiv:hep-ph/9807297
  [hep-ph]}}.

\bibitem{deFlorian:2003qf}
D.~de~Florian and R.~Sassot, ``{Nuclear parton distributions at next-to-leading
  order},'' \href{http://dx.doi.org/10.1103/PhysRevD.69.074028}{{\em Phys.
  Rev.} {\bfseries D69} (2004) 074028},
\href{http://arxiv.org/abs/hep-ph/0311227}{{\ttfamily arXiv:hep-ph/0311227
  [hep-ph]}}.

\end{thebibliography}\endgroup

\newpage
\appendix
\section{The ALICE Collaboration}
\label{app:collab}



\begingroup
\small
\begin{flushleft}
J.~Adam$^\textrm{\scriptsize 39}$,
D.~Adamov\'{a}$^\textrm{\scriptsize 85}$,
M.M.~Aggarwal$^\textrm{\scriptsize 89}$,
G.~Aglieri Rinella$^\textrm{\scriptsize 35}$,
M.~Agnello$^\textrm{\scriptsize 112}$\textsuperscript{,}$^\textrm{\scriptsize 31}$,
N.~Agrawal$^\textrm{\scriptsize 48}$,
Z.~Ahammed$^\textrm{\scriptsize 136}$,
S.~Ahmad$^\textrm{\scriptsize 18}$,
S.U.~Ahn$^\textrm{\scriptsize 69}$,
S.~Aiola$^\textrm{\scriptsize 140}$
% ...
\renewcommand\labelenumi{\textsuperscript{\theenumi}~}

\section*{Affiliation notes}
\renewcommand\theenumi{\roman{enumi}}
\begin{Authlist}
\item \Adef{0}Deceased
\item \Adef{idp1837696}{Also at: Georgia State University, Atlanta, Georgia, United States}
\item \Adef{idp3272992}{Also at: Also at Department of Applied Physics, Aligarh Muslim University, Aligarh, India}
\item \Adef{idp4005328}{Also at: M.V. Lomonosov Moscow State University, D.V. Skobeltsyn Institute of Nuclear, Physics, Moscow, Russia}
\end{Authlist}

\section*{Collaboration Institutes}
\renewcommand\theenumi{\arabic{enumi}~}

$^{1}$A.I. Alikhanyan National Science Laboratory (Yerevan Physics Institute) Foundation, Yerevan, Armenia
\\
$^{2}$Benem\'{e}rita Universidad Aut\'{o}noma de Puebla, Puebla, Mexico
\\
$^{3}$Bogolyubov Institute for Theoretical Physics, Kiev, Ukraine
\\
$^{4}$Bose Institute, Department of Physics 
and Centre for Astroparticle Physics and Space Science (CAPSS), Kolkata, India
\\
$^{5}$Budker Institute for Nuclear Physics, Novosibirsk, Russia
\\
$^{6}$California Polytechnic State University, San Luis Obispo, California, United States
\\
$^{7}$Central China Normal University, Wuhan, China
\\
$^{8}$Centre de Calcul de l'IN2P3, Villeurbanne, Lyon, France
\\
$^{9}$Centro de Aplicaciones Tecnol\'{o}gicas y Desarrollo Nuclear (CEADEN), Havana, Cuba
\\
$^{10}$Centro de Investigaciones Energ\'{e}ticas Medioambientales y Tecnol\'{o}gicas (CIEMAT), Madrid, Spain
\\
$^{11}$Centro de Investigaci\'{o}n y de Estudios Avanzados (CINVESTAV), Mexico City and M\'{e}rida, Mexico
\\
$^{12}$Centro Fermi - Museo Storico della Fisica e Centro Studi e Ricerche ``Enrico Fermi', Rome, Italy
\\
$^{13}$Chicago State University, Chicago, Illinois, United States
\\
$^{14}$China Institute of Atomic Energy, Beijing, China
\\
$^{15}$Commissariat \`{a} l'Energie Atomique, IRFU, Saclay, France
\\
$^{16}$COMSATS Institute of Information Technology (CIIT), Islamabad, Pakistan
\\
$^{17}$Departamento de F\'{\i}sica de Part\'{\i}culas and IGFAE, Universidad de Santiago de Compostela, Santiago de Compostela, Spain
\\
$^{18}$Department of Physics, Aligarh Muslim University, Aligarh, India
\\
$^{19}$Department of Physics, Ohio State University, Columbus, Ohio, United States
\\
$^{20}$Department of Physics, Sejong University, Seoul, South Korea
\\
$^{21}$Department of Physics, University of Oslo, Oslo, Norway
\\
$^{22}$Department of Physics and Technology, University of Bergen, Bergen, Norway
\\
$^{23}$Dipartimento di Fisica dell'Universit\`{a} 'La Sapienza'
and Sezione INFN, Rome, Italy
\\
$^{24}$Dipartimento di Fisica dell'Universit\`{a}
and Sezione INFN, Cagliari, Italy
\\
$^{25}$Dipartimento di Fisica dell'Universit\`{a}
and Sezione INFN, Trieste, Italy
\\
$^{26}$Dipartimento di Fisica dell'Universit\`{a}
and Sezione INFN, Turin, Italy
\\
$^{27}$Dipartimento di Fisica e Astronomia dell'Universit\`{a}
and Sezione INFN, Bologna, Italy
\\
$^{28}$Dipartimento di Fisica e Astronomia dell'Universit\`{a}
and Sezione INFN, Catania, Italy
\\
$^{29}$Dipartimento di Fisica e Astronomia dell'Universit\`{a}
and Sezione INFN, Padova, Italy
\\
$^{30}$Dipartimento di Fisica `E.R.~Caianiello' dell'Universit\`{a}
and Gruppo Collegato INFN, Salerno, Italy
\\
$^{31}$Dipartimento DISAT del Politecnico and Sezione INFN, Turin, Italy
\\
$^{32}$Dipartimento di Scienze e Innovazione Tecnologica dell'Universit\`{a} del Piemonte Orientale and INFN Sezione di Torino, Alessandria, Italy
\\
$^{33}$Dipartimento Interateneo di Fisica `M.~Merlin'
and Sezione INFN, Bari, Italy
\\
$^{34}$Division of Experimental High Energy Physics, University of Lund, Lund, Sweden
\\
$^{35}$European Organization for Nuclear Research (CERN), Geneva, Switzerland
\\
$^{36}$Excellence Cluster Universe, Technische Universit\"{a}t M\"{u}nchen, Munich, Germany
\\
$^{37}$Faculty of Engineering, Bergen University College, Bergen, Norway
\\
$^{38}$Faculty of Mathematics, Physics and Informatics, Comenius University, Bratislava, Slovakia
\\
$^{39}$Faculty of Nuclear Sciences and Physical Engineering, Czech Technical University in Prague, Prague, Czech Republic
\\
$^{40}$Faculty of Science, P.J.~\v{S}af\'{a}rik University, Ko\v{s}ice, Slovakia
\\
$^{41}$Faculty of Technology, Buskerud and Vestfold University College, Tonsberg, Norway
\\
$^{42}$Frankfurt Institute for Advanced Studies, Johann Wolfgang Goethe-Universit\"{a}t Frankfurt, Frankfurt, Germany
\\
$^{43}$Gangneung-Wonju National University, Gangneung, South Korea
\\
$^{44}$Gauhati University, Department of Physics, Guwahati, India
\\
$^{45}$Helmholtz-Institut f\"{u}r Strahlen- und Kernphysik, Rheinische Friedrich-Wilhelms-Universit\"{a}t Bonn, Bonn, Germany
\\
$^{46}$Helsinki Institute of Physics (HIP), Helsinki, Finland
\\
$^{47}$Hiroshima University, Hiroshima, Japan
\\
$^{48}$Indian Institute of Technology Bombay (IIT), Mumbai, India
\\
$^{49}$Indian Institute of Technology Indore, Indore, India
\\
$^{50}$Indonesian Institute of Sciences, Jakarta, Indonesia
\\
$^{51}$Inha University, Incheon, South Korea
\\
$^{52}$Institut de Physique Nucl\'eaire d'Orsay (IPNO), Universit\'e Paris-Sud, CNRS-IN2P3, Orsay, France
\\
$^{53}$Institute for Nuclear Research, Academy of Sciences, Moscow, Russia
\\
$^{54}$Institute for Subatomic Physics of Utrecht University, Utrecht, Netherlands
\\
$^{55}$Institute for Theoretical and Experimental Physics, Moscow, Russia
\\
$^{56}$Institute of Experimental Physics, Slovak Academy of Sciences, Ko\v{s}ice, Slovakia
\\
$^{57}$Institute of Physics, Academy of Sciences of the Czech Republic, Prague, Czech Republic
\\
$^{58}$Institute of Physics, Bhubaneswar, India
\\
$^{59}$Institute of Space Science (ISS), Bucharest, Romania
\\
$^{60}$Institut f\"{u}r Informatik, Johann Wolfgang Goethe-Universit\"{a}t Frankfurt, Frankfurt, Germany
\\
$^{61}$Institut f\"{u}r Kernphysik, Johann Wolfgang Goethe-Universit\"{a}t Frankfurt, Frankfurt, Germany
\\
$^{62}$Institut f\"{u}r Kernphysik, Westf\"{a}lische Wilhelms-Universit\"{a}t M\"{u}nster, M\"{u}nster, Germany
\\
$^{63}$Instituto de Ciencias Nucleares, Universidad Nacional Aut\'{o}noma de M\'{e}xico, Mexico City, Mexico
\\
$^{64}$Instituto de F\'{\i}sica, Universidad Nacional Aut\'{o}noma de M\'{e}xico, Mexico City, Mexico
\\
$^{65}$Institut Pluridisciplinaire Hubert Curien (IPHC), Universit\'{e} de Strasbourg, CNRS-IN2P3, Strasbourg, France
\\
$^{66}$iThemba LABS, National Research Foundation, Somerset West, South Africa
\\
$^{67}$Joint Institute for Nuclear Research (JINR), Dubna, Russia
\\
$^{68}$Konkuk University, Seoul, South Korea
\\
$^{69}$Korea Institute of Science and Technology Information, Daejeon, South Korea
\\
$^{70}$KTO Karatay University, Konya, Turkey
\\
$^{71}$Laboratoire de Physique Corpusculaire (LPC), Clermont Universit\'{e}, Universit\'{e} Blaise Pascal, CNRS--IN2P3, Clermont-Ferrand, France
\\
$^{72}$Laboratoire de Physique Subatomique et de Cosmologie, Universit\'{e} Grenoble-Alpes, CNRS-IN2P3, Grenoble, France
\\
$^{73}$Laboratori Nazionali di Frascati, INFN, Frascati, Italy
\\
$^{74}$Laboratori Nazionali di Legnaro, INFN, Legnaro, Italy
\\
$^{75}$Lawrence Berkeley National Laboratory, Berkeley, California, United States
\\
$^{76}$Moscow Engineering Physics Institute, Moscow, Russia
\\
$^{77}$Nagasaki Institute of Applied Science, Nagasaki, Japan
\\
$^{78}$National Centre for Nuclear Studies, Warsaw, Poland
\\
$^{79}$National Institute for Physics and Nuclear Engineering, Bucharest, Romania
\\
$^{80}$National Institute of Science Education and Research, Bhubaneswar, India
\\
$^{81}$National Research Centre Kurchatov Institute, Moscow, Russia
\\
$^{82}$Niels Bohr Institute, University of Copenhagen, Copenhagen, Denmark
\\
$^{83}$Nikhef, Nationaal instituut voor subatomaire fysica, Amsterdam, Netherlands
\\
$^{84}$Nuclear Physics Group, STFC Daresbury Laboratory, Daresbury, United Kingdom
\\
$^{85}$Nuclear Physics Institute, Academy of Sciences of the Czech Republic, \v{R}e\v{z} u Prahy, Czech Republic
\\
$^{86}$Oak Ridge National Laboratory, Oak Ridge, Tennessee, United States
\\
$^{87}$Petersburg Nuclear Physics Institute, Gatchina, Russia
\\
$^{88}$Physics Department, Creighton University, Omaha, Nebraska, United States
\\
$^{89}$Physics Department, Panjab University, Chandigarh, India
\\
$^{90}$Physics Department, University of Athens, Athens, Greece
\\
$^{91}$Physics Department, University of Cape Town, Cape Town, South Africa
\\
$^{92}$Physics Department, University of Jammu, Jammu, India
\\
$^{93}$Physics Department, University of Rajasthan, Jaipur, India
\\
$^{94}$Physikalisches Institut, Eberhard Karls Universit\"{a}t T\"{u}bingen, T\"{u}bingen, Germany
\\
$^{95}$Physikalisches Institut, Ruprecht-Karls-Universit\"{a}t Heidelberg, Heidelberg, Germany
\\
$^{96}$Physik Department, Technische Universit\"{a}t M\"{u}nchen, Munich, Germany
\\
$^{97}$Purdue University, West Lafayette, Indiana, United States
\\
$^{98}$Pusan National University, Pusan, South Korea
\\
$^{99}$Research Division and ExtreMe Matter Institute EMMI, GSI Helmholtzzentrum f\"ur Schwerionenforschung, Darmstadt, Germany
\\
$^{100}$Rudjer Bo\v{s}kovi\'{c} Institute, Zagreb, Croatia
\\
$^{101}$Russian Federal Nuclear Center (VNIIEF), Sarov, Russia
\\
$^{102}$Saha Institute of Nuclear Physics, Kolkata, India
\\
$^{103}$School of Physics and Astronomy, University of Birmingham, Birmingham, United Kingdom
\\
$^{104}$Secci\'{o}n F\'{\i}sica, Departamento de Ciencias, Pontificia Universidad Cat\'{o}lica del Per\'{u}, Lima, Peru
\\
$^{105}$Sezione INFN, Bari, Italy
\\
$^{106}$Sezione INFN, Bologna, Italy
\\
$^{107}$Sezione INFN, Cagliari, Italy
\\
$^{108}$Sezione INFN, Catania, Italy
\\
$^{109}$Sezione INFN, Padova, Italy
\\
$^{110}$Sezione INFN, Rome, Italy
\\
$^{111}$Sezione INFN, Trieste, Italy
\\
$^{112}$Sezione INFN, Turin, Italy
\\
$^{113}$SSC IHEP of NRC Kurchatov institute, Protvino, Russia
\\
$^{114}$Stefan Meyer Institut f\"{u}r Subatomare Physik (SMI), Vienna, Austria
\\
$^{115}$SUBATECH, Ecole des Mines de Nantes, Universit\'{e} de Nantes, CNRS-IN2P3, Nantes, France
\\
$^{116}$Suranaree University of Technology, Nakhon Ratchasima, Thailand
\\
$^{117}$Technical University of Ko\v{s}ice, Ko\v{s}ice, Slovakia
\\
$^{118}$Technical University of Split FESB, Split, Croatia
\\
$^{119}$The Henryk Niewodniczanski Institute of Nuclear Physics, Polish Academy of Sciences, Cracow, Poland
\\
$^{120}$The University of Texas at Austin, Physics Department, Austin, Texas, United States
\\
$^{121}$Universidad Aut\'{o}noma de Sinaloa, Culiac\'{a}n, Mexico
\\
$^{122}$Universidade de S\~{a}o Paulo (USP), S\~{a}o Paulo, Brazil
\\
$^{123}$Universidade Estadual de Campinas (UNICAMP), Campinas, Brazil
\\
$^{124}$Universidade Federal do ABC, Santo Andre, Brazil
\\
$^{125}$University of Houston, Houston, Texas, United States
\\
$^{126}$University of Jyv\"{a}skyl\"{a}, Jyv\"{a}skyl\"{a}, Finland
\\
$^{127}$University of Liverpool, Liverpool, United Kingdom
\\
$^{128}$University of Tennessee, Knoxville, Tennessee, United States
\\
$^{129}$University of the Witwatersrand, Johannesburg, South Africa
\\
$^{130}$University of Tokyo, Tokyo, Japan
\\
$^{131}$University of Tsukuba, Tsukuba, Japan
\\
$^{132}$University of Zagreb, Zagreb, Croatia
\\
$^{133}$Universit\'{e} de Lyon, Universit\'{e} Lyon 1, CNRS/IN2P3, IPN-Lyon, Villeurbanne, Lyon, France
\\
$^{134}$Universit\`{a} di Brescia, Brescia, Italy
\\
$^{135}$V.~Fock Institute for Physics, St. Petersburg State University, St. Petersburg, Russia
\\
$^{136}$Variable Energy Cyclotron Centre, Kolkata, India
\\
$^{137}$Warsaw University of Technology, Warsaw, Poland
\\
$^{138}$Wayne State University, Detroit, Michigan, United States
\\
$^{139}$Wigner Research Centre for Physics, Hungarian Academy of Sciences, Budapest, Hungary
\\
$^{140}$Yale University, New Haven, Connecticut, United States
\\
$^{141}$Yonsei University, Seoul, South Korea
\\
$^{142}$Zentrum f\"{u}r Technologietransfer und Telekommunikation (ZTT), Fachhochschule Worms, Worms, Germany
\endgroup

\end{document}